\documentclass[12pt,preprint]{aastex}

\usepackage[small]{caption}

\newcommand{\HST}{{\it HST}}
\newcommand{\kms}{\ifmmode {\rm km\ s}^{-1} \else km s$^{-1}$\fi}
\newcommand{\Msun}{\ifmmode {\rm M}_{\odot} \else M$_{\odot}$\fi}
\newcommand{\Lsun}{\ifmmode {\rm L}_{\odot} \else L$_{\odot}$\fi}
\newcommand{\qo}{\ifmmode q_{\rm o} \else $q_{\rm o}$\fi}
\newcommand{\Ho}{\ifmmode H_{\rm o} \else $H_{\rm o}$\fi}
\newcommand{\ho}{\ifmmode h_{\rm o} \else $h_{\rm o}$\fi}

\newcommand{\vFWHM}{\ifmmode v_{\mbox{\tiny FWHM}} \else
                    $v_{\mbox{\tiny FWHM}}$\fi}
\newcommand{\CCF}{\ifmmode F_{\it CCF} \else $F_{\it CCF}$\fi}
\newcommand{\ACF}{\ifmmode F_{\it ACF} \else $F_{\it ACF}$\fi}
\newcommand{\Halpha}{\ifmmode {\rm H}\alpha \else H$\alpha$\fi}
\newcommand{\Hbeta}{\ifmmode {\rm H}\beta \else H$\beta$\fi}
\newcommand{\Hgamma}{\ifmmode {\rm H}\gamma \else H$\gamma$\fi}
\newcommand{\Hdelta}{\ifmmode {\rm H}\delta \else H$\delta$\fi}
\newcommand{\Lya}{\ifmmode {\rm Ly}\alpha \else Ly$\alpha$\fi}
\newcommand{\Lyb}{\ifmmode {\rm Ly}\beta \else Ly$\beta$\fi}
\newcommand{\HeI}{\ifmmode {\rm He}\,{\sc i}\,\lambda5876 \else 
	          He\,{\sc i}\,$\lambda5876$\fi}
\newcommand{\HeII}{\ifmmode {\rm He}\,{\sc ii}\,\lambda4686 \else 
	           He\,{\sc ii}\,$\lambda4686$\fi}

\newcommand{\ciii}{\ifmmode {\rm C}\,{\sc iii} \else C\,{\sc iii}\fi}

\newcommand{\oiii}{\ifmmode {\rm O}\,{\sc iii} \else O\,{\sc iii}\fi}
\newcommand{\ob}{[O\,{\sc iii}]\,$\lambda \lambda 4959,5007$}

\newcommand{\Flamunit}{ergs s$^{-1}$\,cm$^{-2}$\,\AA$^{-1}$}

\newcommand{\rl}{$R_{\rm BLR}$--$L$}
\newcommand{\msig}{$M_{\rm BH}$--$\sigma_{\star}$}
\newcommand{\ml}{$M_{\rm BH}$--$L_{\rm bulge}$}

\shorttitle{Black Hole Masses from Reverberation Mapping}
\shortauthors{}

\received{}
\accepted{}

\begin{document}

\title{Reverberation Mapping Measurements of Black Hole Masses in
Six Local Seyfert Galaxies}

\author{ K.~D.~Denney\altaffilmark{1}, B.~M.~Peterson\altaffilmark{1,2},
         R.~W.~Pogge\altaffilmark{1,2}, A.~Adair\altaffilmark{3},
         D.~W.~Atlee\altaffilmark{1}, K.~Au-Yong\altaffilmark{3},
         M.~C.~Bentz\altaffilmark{1,4,5}, J.~C.~Bird\altaffilmark{1},
         D.~J.~Brokofsky\altaffilmark{6,7}, E.~Chisholm\altaffilmark{3},
         M.~L.~Comins\altaffilmark{1,8}, M.~Dietrich\altaffilmark{1},
         V.~T.~Doroshenko\altaffilmark{9,10,11},
         J.~D.~Eastman\altaffilmark{1}, Y.~S.~Efimov\altaffilmark{10},
         S.~Ewald\altaffilmark{3}, S.~Ferbey\altaffilmark{3},
         C.~M.~Gaskell\altaffilmark{6,12},
         C.~H.~Hedrick\altaffilmark{6,8}, K.~Jackson\altaffilmark{3},
         S.~A.~Klimanov\altaffilmark{10,11},
         E.~S.~Klimek\altaffilmark{6,13}, A.~K.~Kruse\altaffilmark{6,14},
         A.~Lad\'{e}route\altaffilmark{3}, J.~B.~Lamb\altaffilmark{15},
         K.~Leighly\altaffilmark{16}, T.~Minezaki\altaffilmark{17},
         S.~V.~Nazarov\altaffilmark{10,11}, C.~A.~Onken\altaffilmark{18,19},
         E.~A.~Petersen\altaffilmark{6}, P.~Peterson\altaffilmark{20},
         S.~Poindexter\altaffilmark{1}, Y.~Sakata\altaffilmark{21},
         K.~J.~Schlesinger\altaffilmark{1},
         S.~G.~Sergeev\altaffilmark{10,11}, N.~Skolski\altaffilmark{3},
         L.~Stieglitz\altaffilmark{3}, J.~J.~Tobin\altaffilmark{15},
         C.~Unterborn\altaffilmark{1},
         M.~Vestergaard\altaffilmark{22,23},
         A.~E.~Watkins\altaffilmark{6}, L.~C.~Watson\altaffilmark{1},
         and Y.~Yoshii\altaffilmark{17} }

\altaffiltext{1}{Department of Astronomy, 
		The Ohio State University, 
		140 West 18th Avenue, 
		Columbus, OH 43210, USA; 
		denney@astronomy.ohio-state.edu}

\altaffiltext{2}{Center for Cosmology and AstroParticle Physics, 
                 The Ohio State University,
		 191 West Woodruff Avenue, 
		 Columbus, OH 43210, USA}

\altaffiltext{3}{Centre of the Universe,
                 Herzberg Institute of Astrophysics,
		 National Research Council of Canada,
		 5071 West Saanich Road,
		 Victoria, BC  V9E 2E7,
		 Canada}

\altaffiltext{4}{Present address: 
		 Department of Physics and Astronomy,
		 4129 Frederick Reines Hall,
		 University of California at Irvine,
		 Irvine, CA 92697-4575, USA}

\altaffiltext{5}{Hubble Fellow}

\altaffiltext{6}{Department of Physics \& Astronomy, 
		 University of Nebraska, 
		 Lincoln, NE 68588-0111, USA. }

\altaffiltext{7}{Deceased, 2008 September 13}

\altaffiltext{8}{Current address: 
		 Astronomy and Astrophysics Department, 
		 Pennsylvania State University, 
		 525 Davey Laboratory, University Park, PA 16802, USA}

\altaffiltext{9}{Crimean Laboratory of the Sternberg Astronomical Institute, 
	         p/o Nauchny, 98409 Crimea, Ukraine}

\altaffiltext{10}{Crimean Astrophysical Observatory,
		 p/o Nauchny, 98409 Crimea, Ukraine}

\altaffiltext{11}{Isaak Newton Institute of Chile,
	          Crimean Branch, Ukraine}

\altaffiltext{12}{Current address: 
		 Astronomy Department, 
		 University of Texas, 
		 Austin, TX 78712-0259, USA}

\altaffiltext{13}{Current address: 
	          Astronomy Department, MSC 4500,
		  New Mexico State University, 
		  PO BOX 30001, La Cruces, NM 88003-8001, USA}

\altaffiltext{14}{Current address: 
                  Physics Department,
                  University of Wisconsin-Madison,
                  1150 University Avenue,
                  Madison, WI 53706-1390, USA}

\altaffiltext{15}{Department of Astronomy,
		 University of Michigan,
		 500 Church St., 
		 Ann Arbor, MI 48109-1040, USA}

\altaffiltext{16}{Homer L. Dodge Department of Physics and Astronomy,
	    	  The University of Oklahoma,
  		  440 W. Brooks St.,
  		  Norman, OK 73019, USA}
	
\altaffiltext{17}{Institute of Astronomy, 
		 School of Science, 
		 University of Tokyo,
	 	 2-21-1 Osawa, Mitaka, 
		 Tokyo 181-0015, Japan}

\altaffiltext{18}{Plaskett Fellow; Dominion Astrophysical Observatory,
                  Herzberg Institute of Astrophysics, 
		  National Research Council of Canada,
		  5071 West Saanich Road, Victoria, BC V9E 2E7, 
		  Canada}

\altaffiltext{19}{Current address:
                 Mount Stromlo Observatory,
		 Research School of Astronomy \& Astrophysics,
		 The Australian National University,
		 Cotter Road,
		 Weston Creek, ACT 2611,
		 Australia}

\altaffiltext{20}{Ohio University,
		  Department of Physics and Astronomy,
		  Athens, OH 45701-2979, USA}

\altaffiltext{21}{Department of Astronomy, 
		  School of Science, 
		  University of Tokyo,
  		  7-3-1 Hongo, Bunkyo-ku, 
		  Tokyo 113-0013, Japan}

\altaffiltext{22}{Steward Observatory, 
		The University of Arizona, 
		933 North Cherry Avenue, 
         	Tucson, AZ 85721, USA}

\altaffiltext{23}{Dark Cosmology Centre,
                 Niels Bohr Institute,
                 Copenhagan University}

\begin{abstract}

We present the final results from a high sampling rate, multi-month,
spectrophotometric reverberation mapping campaign undertaken to obtain
either new or improved \Hbeta\ reverberation lag measurements for
several relatively low-luminosity AGNs.  We have reliably measured the
time delay between variations in the continuum and \Hbeta\ emission line
in six local Seyfert 1 galaxies.  These measurements are used to
calculate the mass of the supermassive black hole at the center of each
of these AGNs.  We place our results in context to the most current
calibration of the broad-line region (BLR) $R_{\rm BLR}$--$L$
relationship, where our results remove outliers and reduce the scatter
at the low-luminosity end of this relationship.  We also present
velocity-resolved \Hbeta\ time delay measurements for our complete
sample, though the clearest velocity-resolved kinematic signatures have
already been published.
\end{abstract}



\section{INTRODUCTION}

The technique of reverberation mapping \citep{Blandford82,Peterson93}
has been used to directly measure black hole masses in relatively local
broad-line (Type 1) AGNs for over two decades \citep[see compilation
by][]{Peterson04}.  In recent years, these measurements have become
particularly desirable with the increasingly strong evidence (both
observational and theoretical) that there is a connection between
supermassive black hole (BH) growth and galaxy evolution
\citep[e.g.,][]{Silk&Rees98, Kormendy&Gebhardt01, Haring&Rix04,
DiMatteo05, Bennert08, Somerville08, Hopkins09, Shankar09}.  Empirical
relationships have been discovered for both quiescent and active
galaxies that show similar correlations between the central BH and
properties of the bulge of the host galaxy (well outside the
gravitational sphere of influence of the black hole).  Examples include
correlations between the BH mass and total luminosity of stars in the
galactic bulge --- the \ml\ relationship
\citep{Kormendy&Richstone95,Magorrian98, Wandel02,Graham07,Bentz09b} ---
and between BH mass and the bulge stellar velocity dispersion --- the
\msig\ relationship \citep{Ferrarese00, Gebhardt00a, Gebhardt00b,
Ferrarese01,Tremaine02, Onken04, Nelson04}.

The current thrust to better understand this BH-galaxy connection relies
on mass measurements of large samples of black holes in both the local
and distant Universe.  The masses of BHs in distant galaxies can only be
measured indirectly using the scaling relationships mentioned above, as
well as the AGN \rl\ relationship \citep{Kaspi00,Kaspi05,Bentz06a,
Bentz09a}, which provides the capability to estimate BH masses from a
single spectrum of an AGN \citep{Wandel99}.  In order to understand the
evolution of BH and galaxy growth over cosmological times, it is useful
to compare the location of distant galaxies on these relationships with
local samples.  This can only be done by calibrating the local relation
with direct BH mass measurements.

Local masses are measured directly in quiescent galaxies using dynamical
methods \citep[see][for reviews]{Kormendy&Richstone95,
Kormendy&Gebhardt01, Ferrarese05} that rely on resolving the motions of
gas and stars within the sphere of influence of the central BH and are
thus very resolution intensive and only applicable in the nearby
Universe.  Direct measurements can also be made from observations of
megamasers sometimes seen in Type 2 AGNs, but making these observations
relies on a particular viewing angle into the nuclear region of these
galaxies and is thus not applicable to large numbers of objects.  Direct
mass measurements can also be made in Type 1 AGNs using reverberation
mapping, which is a method that relies on time resolution to trace the
light-travel time delay between continuum and broad emission-line flux
variations to measure the characteristic size of the broad line region
(BLR).  Using virial arguments, this size is related to the black hole
mass through the velocity dispersion of the BLR gas, determined from the
broad emission-line width.  Although reverberation mapping is
technically applicable at all redshifts, the reverberation time-delay
scales with the AGN luminosity (i.e., the \rl\ relationship), and this
coupled with time dilation effects make it difficult and particularly
time-consuming to make such measurements out to high redshift
\citep[see][]{Kaspi07}.

The constraints for making direct BH mass measurements at large
distances make the use of the \rl\ relationship particularly attractive
for obtaining even indirect mass estimates at all redshifts for which a
broad-line AGN spectrum can be obtained.  In addition, masses can be
estimated for large samples of objects \citep[e.g.,][]{McLure04,
Kollmeier06, Salviander07, YShen08, Vestergaard08}, facilitating studies
of the BH-galaxy connection and its evolution across cosmic time
\citep[e.g.,][]{Salviander07, Vestergaard&Osmer09}.  However, in order
to reliably apply these relationships to high redshift objects and
determine any evolution in the relationships themselves, local versions
of the relationships need to be well-populated with high-quality data,
so that calibration of these local relationships is secure (i.e.,
observational scatter minimized) and any intrinsic scatter is well
characterized \citep[see, e.g.,][for recent efforts to improve scaling
relation calibration and characterization of intrinsic
scatter]{Bentz06a, Bentz09b, Bentz09a, Graham07, Gultekin09, Woo10}.
Furthermore, systematic uncertainties also need to be understood and
minimized so that the local relations, on which all other related
studies are based, are as robust as possible.  For instance, systematic
uncertainties are present in the direct, dynamical mass measurements of
the BHs in quiescent galaxies due to model-dependencies of the mass
derivation (e.g., \citealt{Gebhardt09} find more than a factor of two
difference in the measured BH mass in M87 when they include a dark
matter halo in their model; see also \citealt{Shen&Gebhardt10} and
\citealt{vandenBosch&deZeeuw10} for more recent model-dependent changes
made to previously measured quiescent black hole masses that change the
masses by similar amounts, i.e., factors of $\sim$2).  On the other
hand, the reverberation-based masses as we present them (measuring
simply the mean BLR radius from the reverberation time-delay) do not
rely on any physical models; instead, the largest systematic uncertainty
comes from the additional zero-point calibration of the mass scale
\citep{Woo10}.  This calibration is needed due to a number of
uncertainties, such as the relationship between the line-of-sight (LOS)
velocity dispersion measured from the broad-line width and the actual
velocity dispersion of the BLR, systematic effects in determining the
effective radius, and the role of non-gravitational forces.

In this work, we present new reverberation-mapping measurements of the
BLR radius and black hole mass for several nearby Seyfert galaxies from
an intensive spectroscopic and photometric monitoring program.  The
goals of this program are (1) to improve the calibration of local
scaling relationships by populating them with not only additional
high-quality measurements, but also replace previous measurements of
either poor quality or that were suspect for one reason or another, and
(2) to take the method of reverberation mapping one step past its
currently successful application of measuring BLR radii and BH masses to
uncover velocity-resolved structure in the reverberation delays from the
\Hbeta\ emission line.  This velocity-resolved analysis is a first step
towards recovering velocity-dependent \Hbeta\ transfer functions, or
``velocity--delay maps'', which describe the response of the
emission-line to an outburst from the ionizing continuum as a function
of LOS velocity and light-travel time-delay \citep[for a tutorial,
see][]{Peterson01,Horne04}.  Creation of velocity--delay maps provides
valuable knowledge of the structure, inclination, and kinematics of the
BLR, which in turn will reduce systematic uncertainties in
reverberation-based black hole mass measurements.

Our monitoring program spanned more than four months, over which primary
spectroscopic observations were obtained nightly (weather permitting)
for the first three months at MDM Observatory.  Supplementary
observations were gathered from other observatories around the world.
Objects in our sample were targeted because (a) they had short enough
expected lags (i.e., low enough luminosity) that we were likely to see
sufficient variability over the course of our $\sim$3--4 month campaign
to securely measure a reverberation time delay, (b) they appeared as
outliers on AGN scaling relationships and/or had large uncertainties
associated with previous results due to suspected undersampling or other
complications, and (c) previous observations demonstrated the potential
for our high sampling-rate observations to uncover a velocity-resolved
line response to the continuum variations.  We also note that some of
the AGNs observed in this program are among the closest AGNs and are
therefore the best candidates for measuring the central black hole
masses by other direct methods such as modeling of stellar or gas
dynamics, which will allow a direct comparison of mass measurements from
multiple independent techniques.  This paper is arranged such that we
present our observations and analysis in Section \ref{S:ObsAnal}, the
black hole mass measurements are described in Section \ref{S:mbh}, any
velocity-resolved structures that we uncovered are presented in Section
\ref{S:velreslags}, and our results are discussed in Section
\ref{S:discussion}.

\section{Observations and Data Analysis}
\label{S:ObsAnal}

Except where noted, data acquisition and analysis practices employed
here follow closely those laid out by \citet{Denney09b} for the first
results from this campaign on NGC\,4051.  The reader is also referred to
similar previous works, such as \citet{Denney06} and \citet{Peterson04},
for additional details and discussions on these practices.  Throughout
this work, we assume the following cosmology: $\Omega_{m}=0.3$,
$\Omega_{\Lambda}=0.70$, and $H_0 = 70$ km sec$^{-1}$ Mpc$^{-1}$.

\subsection{Spectroscopy}

Spectra of the nuclear region of our complete\footnote{We also monitored
MCG\,08-23-067, but because this object did not vary sufficiently during
our campaign, we did not complete a full reduction and analysis of the
data and do not include it as part of our final, complete sample.}
sample (see Table \ref{tab:objlist}) were obtained daily (weather
permitting) over 89 consecutive nights in Spring 2007 with the 1.3 m
McGraw--Hill telescope at MDM Observatory, and supplemental
spectroscopic observations of most targets were obtained with the 2.6 m
Shajn telescope of the Crimean Astrophysical Observatory (CrAO) and/or
the Plaskett 1.8 m telescope at Dominion Astrophysical Observatory (DAO)
to extend the total campaign duration to $\sim$120 nights.  We used the
Boller and Chivens CCD spectrograph at MDM with the 350 grooves/mm
grating (i.e., a dispersion of 1.33\,\AA/pix) to target the
H$\beta\,\lambda 4861$ and [O\,{\sc iii}]\,$\lambda\lambda 4959, 5007$
emission line region of the optical spectrum.  The position angle was
set to $0 \degr$, with a slit width of 5\farcs0 projected on the sky,
resulting in a spectral resolution of 7.6\,\AA\ across this spectral
region.  We acquired the CrAO spectra with the Nasmith spectrograph and
SPEC-10 $1340 \times 100$ pixel CCD.  For these observations a 3\farcs0
slit was utilized, with a $90 \degr$ position angle.  Spectral
wavelength coverage for this data set was from $\sim$3800--6000\,\AA,
with a dispersion of 1.8\,\AA/pix and a spectral resolution of 7.5\,\AA\
near 5100\,\AA. The actual wavelength coverage is slightly greater than
this, but the red and blue edges of the CCD frame are unusable due to
vignetting.  The DAO observations of the \Hbeta\ region were obtained
with the Cassegrain spectrograph and SITe-5 CCD, where the 400
grooves/mm grating results in a dispersion of 1.1\,\AA/pix.  The slit
width was set to 3\farcs0 with a fixed $90 \degr$ position angle.  This
setup resulted in a resolution of 7.9\,\AA\ around the \Hbeta\ spectral
region.  Figure \ref{fig:meanrms} shows the mean and rms spectra of our
sample based on the MDM observations.  Table \ref{tab:specobs} gives
more detailed statistics of the spectroscopic observations obtained for
each target, including number of observations, time span of
observations, spectral resolution, and spectral extraction window.

A relative flux calibration of each set of spectra was performed using
the $\chi^{2}$ goodness of fit estimator algorithm of
\citet{vanGroningen92} to scale relative fluxes to the [\oiii]\,$\lambda
5007$ constant narrow-line flux.  This algorithm not only makes a
multiplicative scaling to account for the night-to-night differences in
flux in this line caused primarily by aperture affects, but it also
makes slight wavelength shifts to correct for zero-point differences in
the wavelength calibration and small resolution corrections to account
for small variations in the line width caused by variable seeing.  The
best-fit calibration is found by minimizing residuals in the difference
spectrum formed between each individual spectrum and the reference
spectrum, which was taken to be the average of the best spectra of each
object (i.e., those obtained under photometric or near-photometric
conditions).  Because of this multiple-component calibration method, the
final, scaled [\oiii]\,$\lambda 5007$ line flux in each spectrum is not
exactly the same as the reference spectrum.  Instead, there is a small
standard deviation in the mean line flux due to differences in data
quality that averages $\sim$1.2\% across our sample.

\subsection{Photometry}

In addition to spectral observations, we obtained supplemental $V$-band
photometry from the 2.0 m Multicolor Active Galactic NUclei Monitoring
(MAGNUM) telescope at the Haleakala Observatories in Hawaii, the 70 cm
telescope of the CrAO, and the 0.4 m telescope of the University of
Nebraska.  The number of observations obtained from each telescope and
the time span over which observations were made of each target are given
in Table \ref{tab:photobs}.

The MAGNUM observations were made with the multicolor imaging photometer
(MIP) as described by \citet{Kobayashi98a,Kobayashi98b},
\citet{Yoshii02}, and \citet{Kobayashi04}.  Photometric fluxes were
measured within an aperture with radius 8\farcs3.  Reduction of these
observations was similar to that described for other sources by
\citet{Minezaki04} and \citet{Suganuma06}, except the host-galaxy
contribution to the flux within the aperture was not subtracted and the
filter color term was not corrected because these photometric data were
later scaled to the MDM continuum light curves (as described below).
Also, minor corrections (of order 0.01 mag or less) due to the seeing
dependence of the host-galaxy flux were ignored.

The CrAO photometric observations were collected with the AP7p CCD
mounted at the prime focus of the 70 cm telescope ($f=282$ cm).  In this
setup, the $512 \times 512$ pixels of the CCD field projects to a
15\arcmin\ $\times$ 15\arcmin\ field of view.  Photometric fluxes were
measured within an aperture diameter of 15\farcs0.  For further details
of the CrAO $V$-band observations and reduction, see the similar
analysis described by \citet{Sergeev05}.

The University of Nebraska observations were conducted by taking and
separately measuring a large number of one-minute images ($\sim$20).
Details of the observing and reduction procedure are as described by
\citet{Klimek04}.  Comparison star magnitudes were calibrated following
\citet{Doroshenko05a,Doroshenko05b} and \citet{Chonis08}.  To minimize
the effects of variations in the image quality, fluxes were measured
through an aperture of radius 8\farcs0.  The errors given for each
night are the errors in the means.

\subsection{Light Curves}
\label{S:lightcurves}

Except where noted below for individual objects, continuum and \Hbeta\
light curves were created as followed.  Continuum light curves for each
object were made with the $V$-band photometric observations and the
average continuum flux density measured from spectroscopic observations
over the spectral ranges listed in Table \ref{tab:specobs} (i.e., rest
frame $\sim$5100\,\AA).  Continuum light curves from each source were
scaled to the same flux scale following the procedure described by
\citet{Denney09b}.  Figure \ref{fig:lcsep} (top panels) shows these
merged light curves, where measurements from each different observatory
are shown by the different symbols described in the figure caption.

Light curves of the \Hbeta\ flux were made by integrating the line flux
above a linearly interpolated continuum, locally defined by regions just
blueward and redward of the \Hbeta\ emission line.  The \Hbeta\ emission
line was defined between the observed frame wavelength ranges given for
each object in Table \ref{tab:specobs}.  The \Hbeta\ light curves formed
from each separate spectroscopic data set (i.e., MDM, CrAO, and DAO)
were placed on the same flux scale (i.e., that of the MDM observations)
by again following the scaling procedures described by
\citet{Denney09b}.  An additional flux calibration step was used for
NGC\,3516, however, because it has a particularly extended [\oiii]
narrow-line emission region.  In an attempt to decrease the
uncertainties in our relative flux calibration from slit losses of this
extended emission, we made an additional correction to each MDM \Hbeta\
flux measurement to account for possible differences in the observed
[\oiii]\,$\lambda 5007$ flux due to seeing effects.  To measure the
expected differences in [\oiii]\,$\lambda 5007$ flux entering the slit
as a result of changes in the nightly seeing, we followed the procedure
of \citet{Wanders92}, using their artificially seeing-degraded
narrow-band image of the [\oiii]\,$\lambda 5007$ emission from the
nuclear region of NGC\,3516 (details regarding the narrow-band data are
described by Wanders et al.).  Using the differences in measured flux,
we scaled our MDM flux measurements accordingly.  We could only do this
for the MDM measurements, since we do not have accurate seeing estimates
for the CrAO and DAO data sets.  Because of our deliberately large
aperture (see Table \ref{tab:specobs}, Column 8), the effect was not
appreciable for most observations, and there is no indication that our
inability to complete the same analysis for the CrAO and DAO data had
any measurable effect on the subsequent time-series analysis.  The lower
panels of Figure \ref{fig:lcsep} show the \Hbeta\ light curves for each
object after merging the separate data sets into a single \Hbeta\ light
curve.

Before completing the time-series analysis, the light curves shown in
Figure \ref{fig:lcsep} were modified in the following ways:
\begin{enumerate}

\item An absolute flux calibration was applied to both continuum and
\Hbeta\ light curves by scaling to the absolute flux of the
[\oiii]\,$\lambda 5007$ emission line given for each object in Column 3
of Table \ref{tab:constspecprop}.  For objects in which there was not a
previously reported absolute flux, we calculated one from the average
line flux measured from only those observations obtained at MDM under
photometric conditions.

\item The host galaxy starlight contribution to the continuum flux was
subtracted.  This contribution, listed for each target in Column 5 of
Table \ref{tab:constspecprop}, was determined using the methods of
\citet{Bentz09a} for all objects except Mrk\,290, which had not been
targeted for reverberation mapping prior to our observing
campaign\footnote{The 2008 LAMP campaign \citep{Bentz09c} subsequently
monitored Mrk\,290, and it is currently being targeted for \HST\
observations (GO 11662, PI Bentz) to measure its host starlight
contribution, but the observations have not yet been completed.}.  For
Mrk\,290, we use an estimate made from the spectral decomposition
(following decomposition method ``B'' described by \citealt{Denney09a})
of an independent spectrum taken at MDM with nearly the same setup as
our campaign observations but covering optical wavelengths from
3500--7150\,\AA\ with a 1\farcs5 slit.  This value is only a lower
limit, however, since this slit width was smaller than that of our
campaign observations (i.e., 5\farcs0).

\item We ``detrended'' any light curves in which we detected long-term
secular variability over the duration of the campaign that is not
associated with reverberation variations (\citealt{Welsh99}; see also
\citealt{Sergeev07}, who show that there is little correlation between
long-term continuum variability and \Hbeta\ line properties,
demonstrating the independence of this variability on reverberation
processes).  Detrending is important because if the time series contains
long-term trends (i.e., compared to reverberation timescales), the flux
measurements are not randomly distributed about the mean and are, thus,
highly correlated on these long timescales.  These long time scale
correlations then dominate the results of the cross correlation analysis
that determines the time delay, biasing the desired correlation due to
reverberation.  \citet{Welsh99} strongly recommends removing these
low-frequency trends with low order polynomials (a linear fit at the
very least) to improve the reliability of cross correlation lag
determinations.  We took a conservative approach and only linearly
detrended light curves in which there was evidence for secular
variability and for which the cross correlation analysis was improved
upon detrending: both light curves from Mrk\,290, the \Hbeta\ light
curve from Mrk\,817, and the continuum light curve from NGC\,3227 (see
Section \ref{S:lagresults} for further discussion).  These fits are
shown in Figure \ref{fig:lcsep} for each of these respective light
curves.  It was unnecessary to detrend all light curves, as no
improvement in the cross correlation analysis would result from
detrending light curves that already have a relatively flat mean flux.
Also, it is not surprising for associated continuum and line light
curves to exhibit different long-term secular trends, since the
relationship between the measured continuum and the ionizing continuum
responsible for producing the emission lines may not be a linear one
\citep{Peterson02}, and the exact response of the line depends on the
detailed structure and dynamics of the BLR.

\item We excluded the points from the Mrk\,817 light curve with
JD$<$2454200 because (1) there is a large gap in the data between these
points and the rest of the light curve, and (2) there is little to no
coherent variability pattern seen here (i.e., the continuum is
relatively flat and noisy, and the \Hbeta\ fluxes are particularly noisy
and are of otherwise little use, given there are no continuum points at
earlier times).

\end{enumerate}

Tabulated continuum and \Hbeta\ fluxes for all objects, except for
NGC\,4051 which were previously reported by \citet{Denney09b}, are given
in Tables \ref{tab:contflux} and \ref{tab:hbetaflux}, respectively.
Values listed represent the flux of each observation after completing
all flux calibrations described above (i.e., absolute flux calibration
based on the [\oiii]\,$\lambda 5007$ emission-line flux and host galaxy
starlight subtraction), but before detrending, since this results in an
arbitrary flux scale normalized to 1.0.  The final calibrated light
curves used for the subsequent time-series analysis are shown for each
object in the left panels of Figure \ref{fig:lccrosscorr}.  Statistical
parameters describing these calibrated light curves (again, before
detrending) are given in Table \ref{tab:lcstats}, where Column (1) lists
each object.  Columns (2) and (3) are mean and median sampling
intervals, respectively, between data points in the continuum light
curves. The mean continuum flux is shown in column (4), while column (5)
gives the excess variance, calculated as

\begin{equation}
F_{\rm var} = \frac{\sqrt{\sigma^2 - \delta^2}}{\langle f \rangle}
\end{equation}

\noindent where $\sigma^2$ is the variance of the observed fluxes,
$\delta^2$ is their mean square uncertainty, and $\langle f \rangle$ is
the mean of the observed fluxes.  Column (6) is the ratio of the maximum
to minimum flux in the continuum light curves.  Columns (7--11) display
the same quantities as Columns (2--6) but for the \Hbeta\ light curves.

\subsection{Time-Series Analysis}
\label{S:lagresults}

We performed a cross correlation analysis to evaluate the mean
light-travel time delay, or lag, between the continuum and \Hbeta\
emission line flux variations.  We primarily employed an interpolation
scheme (\citealt{Gaskell86,Gaskell87}, with the modifications of
\citealt{White94}).  Using this method, we first interpolate (with an
interval equal to roughly half the median data spacing, i.e., $\sim$0.5
day) between points in the emission-line light curve before cross
correlating it with the original continuum light curve, calculating
cross correlation coefficients, $r$, for many potential lag values (both
positive and negative).  We then average these cross correlation
coefficients with those measured by imposing the same set of possible
lag values in the case where we cross correlate an interpolated
continuum light curve with the original emission-line light curve.  This
gives us a distribution of average cross correlation coefficients as a
function of possible lags, known as the cross correlation function
(CCF).  We checked the results from this method with the discrete
correlation method of \citet{Edelson88}, also employing the
modifications of \citet{White94}, but we do not show these results here,
since they are consistent with our primary cross correlation method, and
provide no additional information.

The right panels of Figure \ref{fig:lccrosscorr} show the adopted cross
correlation results for each object (i.e., after detrending selected
light curves; see below for a discussion of the effect of detrending on
this analysis).  Here, the auto-correlation function (ACF), computed by
cross correlating the continuum with itself, is shown in the top right
panel for each object, and the CCF computed by cross correlating the
\Hbeta\ light curve with that of the continuum, is shown in the bottom
right.  Because the CCF is a convolution of the transfer function with
the ACF, it is instructive to compare the two distributions, as the lag
measured through this type of cross correlation analysis will depend not
only on the delay map, but also on characteristic time scales of the
continuum variations \citep[see, e.g.,][]{Netzer90b}.  We characterize
the time delay between the continuum and emission-line variations by the
parameter $\tau_{\rm cent}$, the centroid of the CCF based on all points
with $r\geq 0.8r_{\rm max}$, as well at the lag corresponding to the
peak in the CCF at $r=r_{\rm max}$, $\tau_{\rm peak}$.  Time
dilation-corrected values of $\tau_{\rm cent}$ and $\tau_{\rm peak}$
were determined for each object using the redshifts listed in Table
\ref{tab:objlist}, i.e., $\tau_{\rm rest} = \tau_{\rm obs}/(1+z)$, and
are given in Table \ref{tab:results}.  Uncertainties in both lag
determinations are computed via model-independent Monte-Carlo
simulations that employ the bootstrap method of \citet{Peterson98}, with
the additional modifications of \citet{Peterson04}.

Visual inspection of the CCFs of selected objects before and after
detrending was made to determine if detrending these light curves was
warranted.  Based on the combined properties of the light curves shown
in Figure \ref{fig:lcsep} (whether or not an overall slope appeared in
the flux across the extent of our campaign) and the CCFs, shown in
Figure \ref{fig:detrCCFs} for Mrk\,290, Mrk\,817, and NGC\,3227 before
and after detrending, we ultimately decided to adopt the detrending for
the following reasons listed for each object:

\medskip
\noindent{\it{Mrk\,290 ---}} The top panels of Figure \ref{fig:detrCCFs}
show that before detrending (left), the peak of the CCF is broader than
the detrended peak (right) and is blended with an aliased peak at
$\sim$30 days.  Since the reverberation lag is clearly seen in the
Mrk\,290 light curves in Figures \ref{fig:lcsep} and
\ref{fig:lccrosscorr} and the peak of highest significance is the same
both before and after detrending, the presence of this alias only acts
to decrease the precision of our lag measurements.  While $\tau_{\rm
cent}$ is roughly one day smaller after detrending (a difference less
than even the measured uncertainty) due to the reduced significance of
the aliased peak at $\sim$30 days by a factor of almost 10, the
detrended CCF is narrower and the measured lags more precise, so we
adopt the detrended measurements.

\medskip
\noindent{\it{Mrk\,817 ---}} The middle panels of Figure
\ref{fig:detrCCFs} show the original (left) and detrended (right) CCFs
from the analysis of Mrk\,817.  The choice to detrend was marginal in
this case.  The process resulted in a larger observed lag ($\tau_{\rm
cent} = 14.48$ days versus $\tau_{\rm cent} = 11.93$) after detrending,
contrary to the typical expectation that lags will be underestimated
after detrending (since the process removes low frequency variability).
We adopt the detrended results because the resulting CCF is narrower,
particularly with respect to lags $\lesssim 0.0$ days, and the resulting
lag measurement is more consistent with past results that we hold to be
reliable (see Section \ref{S:compare2previous}).

\medskip
\noindent{\it{NGC\,3227 ---}} The bottom panels of Figure
\ref{fig:detrCCFs} show the original (left) and detrended (right) CCFs
from the analysis of NGC\,3227.  Here is it obvious that not detrending
the light curves results in a non-physical measurement of the lag at
$\sim$ -33 days with a broad peak (due to aliasing effects between the
features with the highest flux in each of the original continuum and
\Hbeta\ light curves).  While the physical peak (i.e, with positive lag,
as seen and measured from the detrended CCF) is present, every lag is of
low significance, i.e., $r\lesssim$0.4.  After detrending, the CCF peak
at negative lags is still present, however the 'true' reverberation
signal at a lag of $\sim$4 days is rightfully more significant.

\section{Black Hole Masses}
\label{S:mbh}

We assume that the motions of the BLR are dominated by the gravity
of the central black hole so that the mass of the black hole can be
defined by

\begin{equation}
M_{\rm BH} = \frac{f c \tau (\Delta V)^2}{G}.
\end{equation}

\noindent Here, $\tau$ is the measured emission-line time delay, so that
c$\tau$ represents the BLR radius, and $\Delta V$ is the BLR velocity
dispersion.  The dimensionless factor $f$ depends on the structure,
kinematics, and inclination of the BLR, and we adopt the value of
\citet{Onken04}, $f=5.5\pm1.4$, determined empirically by adjusting the
zero-point of the reverberation-based masses to scale the AGN $M_{\rm
BH}$--$\sigma_{\star}$ relationship to that of quiescent galaxies.

\begin{figure}
\figurenum{1}
\epsscale{1}
\plotone{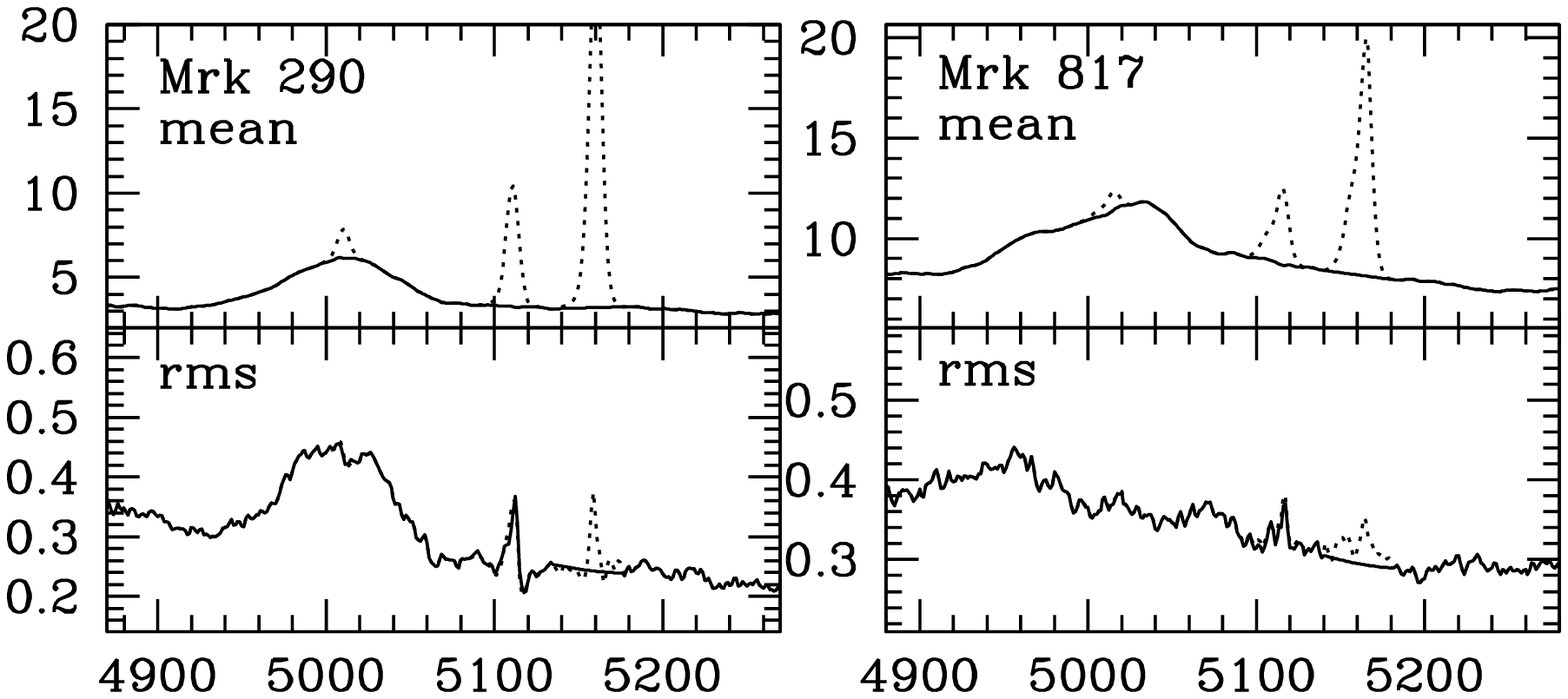}
\plotone{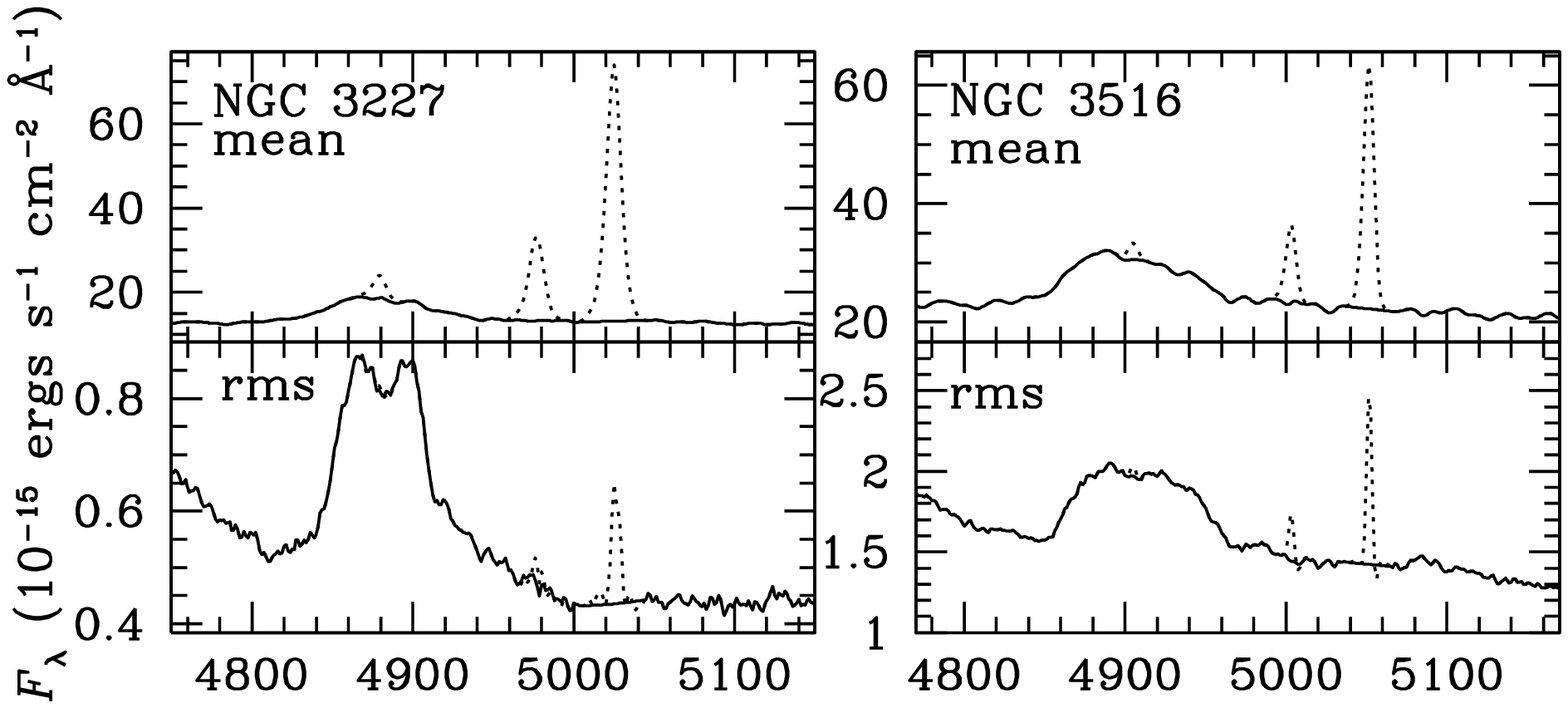}
\plotone{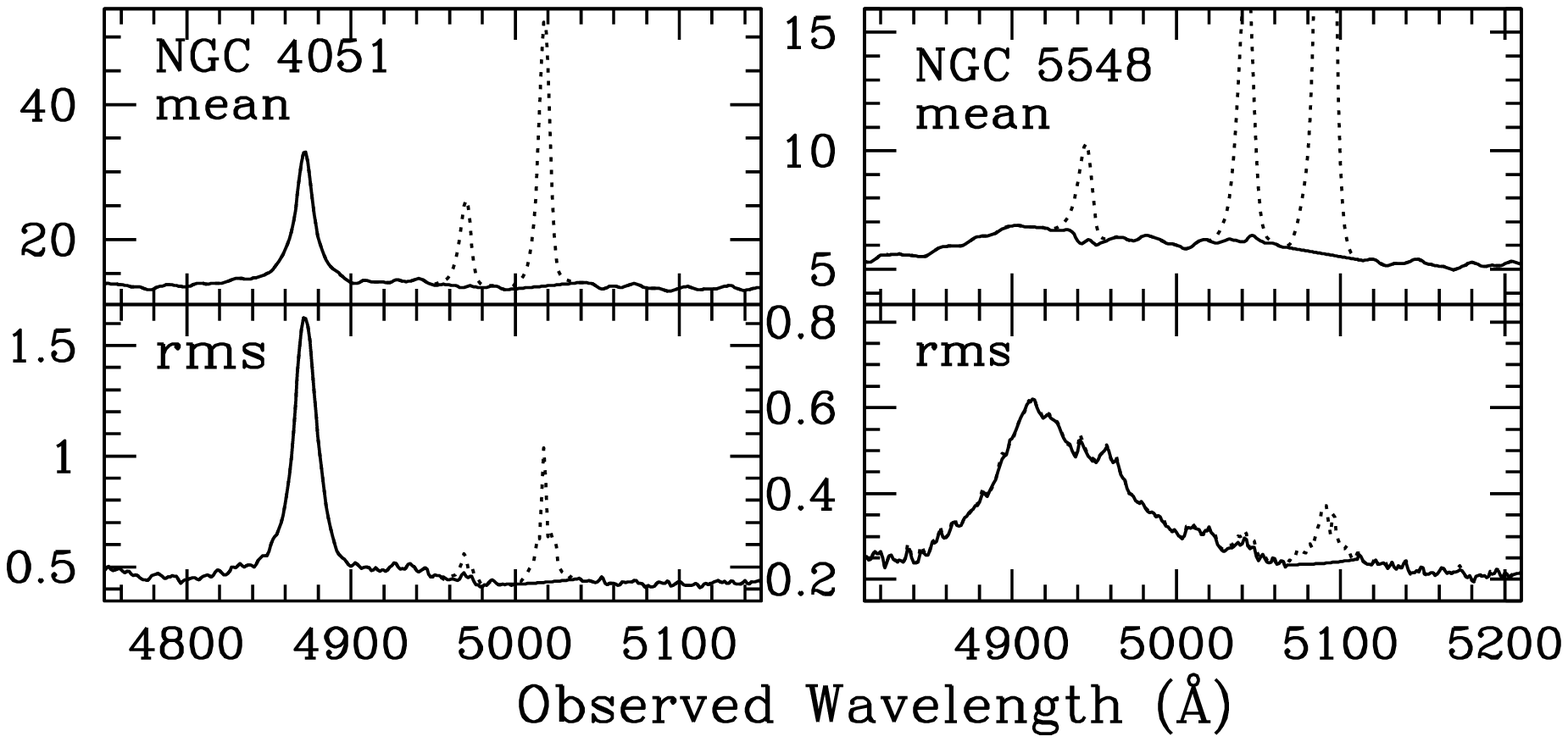}

\caption{Mean and rms (variable emission) spectra from MDM observations.
The solid lines show the narrow-line subtracted spectra, while the
dotted lines show the narrow-line component of \Hbeta\ and the \ob\
narrow emission lines and rms residuals.}

\label{fig:meanrms}
\end{figure}

An estimate of the BLR velocity dispersion is made from the width of the
Doppler-broadened \Hbeta\ emission line.  This line width is commonly
characterized by either the FWHM or the line dispersion, i.e., the
second moment of the line profile.  Table \ref{tab:results} gives both
FWHM and line dispersion, $\sigma_{\rm line}$, measurements from the rms
spectra of all objects except Mrk\,817, in which the rms profile was not
well defined (see Figure \ref{fig:meanrms}), and thus we measured the
width from the mean spectrum.  All widths and their uncertainties were
measured employing methods described in detail by \citet{Peterson04}.
We removed the narrow-line \ob\ emission and the narrow-line component
of \Hbeta\ from all objects before these line widths were measured
(except for NGC\,4051, where this component could not be reliably
isolated due to the line profile shape and, in any case, does not affect
our rms line width measurements; see \citealt{Denney09b}).  Flux
contributions from the narrow-line component will not contaminate the
line widths measured in the rms spectrum (i.e., the narrow-line
component does not vary in response to the ionizing continuum on
reverberation timescales), so removal of this component was generally
unnecessary for most objects in our sample; however, we do so for all
objects anyway to check the accuracy of our \Hbeta\ to
\oiii\,$\lambda$5007 line ratio determinations (Table
\ref{tab:constspecprop}, column 4) by looking for any significant
residual narrow-line emission in the rms spectra of
Figure~\ref{fig:meanrms}.  The exception to this is for Mrk\,817: since
we measured the width in the mean spectrum, it was necessary to remove
the narrow-line before measuring the line widths because the narrow-line
component will bias (i.e., underestimate) line widths measured in the
mean spectrum or in any single-epoch spectrum \citep[see][]{Denney09a}.
Also, for the width measurements in two cases, Mrk\,290 and NGC\,3227,
we narrowed the line boundaries to 4935--5064\,\AA\ and 4810--4942\,\AA,
respectively, compared to what was used for the flux measurements, since
the rms line profiles of these objects were clearly narrower than their
mean profiles (the rms profile is often narrower than the mean profile,
which is not surprising, given that likely not all flux seen in the mean
spectrum varies in response to the continuum; see, e.g.,
\citealt{Korista&Goad04}).

Black hole masses for all objects, calculated from equation (2), are
listed in Table \ref{tab:results} and were calculated using $\tau_{\rm
cent}$, for the time delay, $\tau$, and the quoted line dispersion,
$\sigma_{\rm line}$, for the emission-line width, $\Delta V$.  This
combination of measurements for the line width and reverberation lag is
not only appropriate because it is the combination used by
\citet{Onken04} to determine the value of the scale factor, $f$, that we
adopt here, but also because \citet{Peterson04} show that this
combination also results in the strongest virial relation between line
width and BLR radius, i.e., $R\sim \Delta V^{-0.5}$.  The exception to
this prescription for the black hole mass calculation is Mrk\,817, which
has a poorly defined, triple-peaked rms line profile.  Because the rms
profile is weak and poorly-defined, we measure the line widths from the
mean spectrum and use the \citet{Collin06} calibration of the scale
factor determined for the line dispersion measured from the mean
spectrum, $f=3.85$.  Statistical and observational uncertainties have
been included in these mass measurements, but intrinsic uncertainties
from sources such as unknown BLR inclination cannot be accurately
ascertained.  We also note here that there has been some debate in the
literature as to the importance of radiation pressure on black hole
masses calculated using virial assumptions, since the outward radiation
force has the same radial dependence as gravity \citep[see][]{Marconi08,
Netzer09, Marconi09}.  As there is not yet conclusive evidence
suggesting a radiation-pressure correction is important for the
relatively low Eddington ratio objects we present here, we do not make
this correction, but a radiation-pressure corrected mass can be computed
from the observables given in Table \ref{tab:results} and the formulae
provided by \citet{Marconi08}.

\section{Velocity-Resolved Reverberation Lags}
\label{S:velreslags}

The primary cross correlation analysis presented above was intended to
measure the average time delay across the full extent of the BLR from
which to ascertain the mean, or ``characteristic,'' radius of the
\Hbeta-emitting region of the BLR to use for calculating black hole
masses.  For this reason, we utilized the full line flux from which to
measure the reverberation signal. However, the BLR is an extended
region, and therefore, the light-travel time for the ionizing continuum
to reach different volume elements within the BLR will vary across the
extent of the emitting region.  The expectation is then that the
responding BLR gas variations will lag the continuum variations on
slightly different time scales as a function of the line of sight
velocity.  Measuring and mapping these slight differences in the BLR
response time across velocity space recovers the transfer function,
which is easily visualized as a velocity--delay map
\citep[see][]{Horne04}.  Recovering an unambiguous velocity--delay map
is a continuing goal of reverberation mapping analyses, as the
construction and analysis of such a map is our best hope, with current
technology, of gaining insight into the geometry and kinematics of the
BLR.

The construction and analysis of full two-dimensional velocity--delay
maps is beyond the scope of this work and remains the focus of future
research.  However, we do present a more simple reconstruction of the
velocity-dependent reverberation signal, observed across the \Hbeta\
emission line region when we divide the line flux into eight
velocity-space bins of equal flux.  These results for NGC\,4051,
NGC\,3516, NGC\,3227, and NGC\,5548 have been previously published
\citep{Denney09b,Denney09c} but are included again here for
completeness.  Line boundaries are the same as those used in the full
line analysis, except where noted in Table \ref{tab:specobs}.  In these
cases the narrowed boundaries given above for Mrk\,290 were used, and a
discussion of the difference in boundary choices for the other objects
is presented by \citet{Denney09c}.  Light curves were created from
measurements of the integrated \Hbeta\ flux in each bin and then cross
correlated with the continuum light curve following the same procedures
described above.  Figure \ref{fig:reslags} shows the results of this
analysis for all objects, where the top panel shows the division of each
rms \Hbeta\ line profile into the eight velocity bins, and the bottom
panels shows the lag measurements and uncertainties for each of these
bins.  Error bars in the velocity direction represent the bin width.  We
see a variety of velocity-resolved responses that we discuss in further
detail below.

\section{Discussion}
\label{S:discussion}

\subsection{Comparison with Previous Results}
\label{S:compare2previous}

Some of the objects in this campaign were targeted, at least in part,
because they have previously appeared as outliers on AGN scaling
relationships, in particular, the \rl\ relationship.  As such, all
objects except Mrk\,290 have previous reverberation results, several of
which were suspect for one reason or another and warranted
re-observation.  Based on the outcomes of the current analysis, we will
group our results into three categories: (1) new measurements for an
object never before targeted, i.e., Mrk\,290, (2) replacement
measurements for objects that had uncertain results (typically due to
undersampling) and for which our results completely replace any previous
measurements of the \Hbeta\ reverberation lag, i.e., NGC\,3227,
NGC\,3516, and NGC\,4051, and (3) additional measurements of objects for
which we already trust the previous lag measurements, i.e., NGC\,5548
and Mrk\,817.  In this context, we can compare our new results to
previously published results.

\subsubsection{New Measurements} 

At the time of our campaign (first half of 2007), reverberation mapping
had never before targeted Mrk\,290.  However, in 2008 LAMP also
monitored Mrk\,290 for a reverberation analysis \citep[see][]{Bentz09c},
although they were unable to recover an unambiguous reverberation lag
measurement from their data because Mrk\,290 exhibited little
variability during their campaign.  Therefore, the results we present
here are the only reverberation measurements of this object.

\subsubsection{Replacement Measurements} 

Our current measurements of NGC\,3227, NGC\,3516, and NGC\,4051 should
completely supersede previous results measuring a reverberation radius
based on \Hbeta\ and the black hole mass.  A thorough comparison between
our new measurement of the BLR radius of NGC\,4051 and that from past
studies is discussed by \citet{Denney09b}, and the reader is referred to
this work for details.  However, the main conclusion of that comparison
is that the light curves from which previous measurements of the lag
were made \citep[e.g.,][]{Peterson00b} were undersampled, leading to an
overestimate of the lag.  Our current study remedied this problem with a
much higher sampling rate, routinely obtaining more than one observation
per day.

Previous reverberation lag measurements of the \Hbeta-emitting region in
NGC\,3227 \citep{Salamanca94, Winge95, Onken03} were reanalyzed by
\citet{Peterson04}.  The \Hbeta\ light curves of \citet{Salamanca94}
from a Lovers of Active Galaxies (LAG) campaign were undersampled, and
they do not even attempt to measure a time delay from them.
\citet{Winge95} report an \Hbeta\ lag of $18 \pm 5$ days from
observations taken during a period in which the optical luminosity was
only $\sim$0.3 dex larger than our current observations (i.e., a change
in radius of $\sim$40\% is expected from such a change in luminosity,
based on a \rl\ relationship slope of $\sim$0.5).  However, their
average and median sampling intervals were $\sim$6 and four days,
respectively, which is marginally sampled compared to what is needed for
this low luminosity source.  These early reverberation campaigns did not
have the benefit of the predictive power that we currently have with the
\rl\ relationship to use for planning campaign observations; i.e., these
campaigns were fundamentally exploratory.  A reanalysis of the LAG
consortium data presented by \citet{Salamanca94} was conducted by
\citet{Onken03} using the \citet{vanGroningen92} algorithm to reduce
uncertainties in the relative flux calibration of the spectra.  Onken et
al.\ found an \Hbeta\ lag of $\tau_{\rm cent} = 12.0^{+26.7}_{-9.1}$
days, consistent with the results of \citet{Winge95}.  Later,
\citet{Peterson04} also re-analyzed the CTIO data presented by
\citet{Winge95} with the \citet{vanGroningen92} algorithm and further
re-examined the LAG data rescaled by \citet{Onken03}.  This reanalysis
resulted in some improvement in the \Hbeta\ lag determinations and
uncertainties, i.e., smaller overall lags, however, the reanalyzed
values still had large uncertainties, resulting in a measurement
consistent with zero lag: $\tau_{\rm cent} = 8.2^{+5.1}_{-8.4}$ days and
$\tau_{\rm cent} = 5.4^{+14.1}_{-8.7}$ days for the CTIO and LAG data
sets, respectively \citep{Peterson04}.  It is clear that our new
measurement of the \Hbeta\ lag in NGC\,3227 of $\tau_{\rm cent} =
3.75^{+0.76}_{-0.82}$ days should supersede these past results.

Likewise, the previous reverberation data for NGC\,3516 also came from a
LAG consortium campaign, also with a sampling interval of $\sim$4 days
\citep{Wanders93}.  Since the lag for this object was at least larger
than the sampling rate, the undersampling was not as severe a handicap
as for other objects in our sample, such as NGC\,4051 and NGC\,3227.
Thus, reanalysis of the LAG data first by \citet{Onken03} and then by
\citet{Peterson04} measure lags of $\tau_{\rm cent} = 7.3^{+5.4}_{-2.5}$
days and $\tau_{\rm cent} = 6.7^{+6.8}_{-3.8}$ days, respectively, that
are consistent with the original analysis by Wanders et al., who measure
the peak \Hbeta\ lag to be $7 \pm 3$ days, with the centroid of the CCF
yielding a radius of 11 light days.  All of these centroid measurements
are consistent with our new measurement of $\tau_{\rm cent} =
11.68^{+1.02}_{-1.53}$ days. Also, the LAG spectra were obtained through
a narrow (2\farcs0) slit; as the narrow-line region in this object is
partially resolved, it was necessary to make seeing-dependent
corrections to the continuum and emission-line measurements
\citep{Wanders92} that are both large and uncertain.  For our new
measurements, the aperture corrections are small and have a negligible
effect on the final results; the seeing-corrected and uncorrected fluxes
differ by, on average, $0.09\pm0.05\%$, which is smaller than the
standard deviation of our relative flux scaling of 1.6\% for NGC\,3516.
Clearly, our new observations with an approximately daily sampling rate
show great improvement over past campaigns, for these objects, and the
results presented here should supersede past values of the \Hbeta\ lag
measured for NGC\,3227, NGC\,3516, and NGC\,4051.

\subsubsection{Additional Measurements} 

The goals of this campaign were not only to re-observe outliers or
objects with highly uncertain lag measurements but also to explore the
possibility of uncovering velocity-resolved kinematic signatures and
eventually reconstruct velocity--delay maps.  Therefore, we also
monitored two objects, NGC\,5548 and Mrk\,817, for which previous
reverberation mapping results are solid, and lags measured from this
campaign are simply to be considered additional measurements of the BLR
radius.  Reasons for making repeat reverberation measurements of AGNs
include (1) exploring the radius-luminosity relationship in a single
source, (2) checking the repeatability of the mass measurements for AGNs
at different times, in different luminosity states, and with different
line profiles, and (3) testing different characterizations of the line
width (i.e., determining what line width measure leads to the most
repeatable mass value).  The mean lag and black hole mass results
presented here for NGC\,5548 are consistent with past results, taking
into account the luminosity state of NGC\,5548 during our campaign
compared with other campaigns (i.e., NGC\,5548 has been in a low
luminosity state for the past several years, but the measured lags have
been consistently smaller, as expected for this low state; also see
\citealt{Bentz07,Bentz09c}).

We also monitored Mrk\,817, which is the highest luminosity object in
our present sample.  Previous measurements of the \Hbeta\ radius were
made by \citet{Peterson98} from an eight-year campaign to monitor nine
Seyfert 1 galaxies.  From this campaign, they separately measured the
lag from three different observing seasons.  The reanalysis of this data
by \citet{Peterson04} resulted in rest-frame $\tau_{\rm cent}$
measurements of $19.0^{+3.9}_{-3.7}$, $15.3^{+3.7}_{-3.5}$, and
$33.6^{+6.5}_{-7.6}$ days.  \citet{Bentz09a} calculate a weighted
average of log\,$\tau_{\rm cent}$ from these three measurements of
(converted back to linear space) $\langle \tau_{\rm cent}\rangle_{\rm
wt}=21.8^{+2.4}_{-3.0}$ days at an average luminosity of $\langle {\rm
log}\,L_{5100}\rangle_{\rm wt}=43.64\pm0.03$ to use in calibrating the
\rl\ relationship.  The luminosity of Mrk\,817 during our campaign was
only about 0.1 dex higher than the weighted average luminosity quoted by
Bentz et al., and our measured lag of $\tau_{\rm cent} =
14.04^{+3.41}_{-3.47}$ days is highly consistent with the shortest lag
of Peterson et al.\ and marginally consistent with the 19.0 day lag and
the weighted average.  Furthermore, the virial mass that we measure (see
Column 8 of Table \ref{tab:results}) is also consistent with those given
by \citet{Peterson04}.  Unfortunately, we were not able to improve on
the uncertainties associated with these measurements, as our \Hbeta\
light curve for this object was rather noisy (see Figures
\ref{fig:lcsep} and \ref{fig:lccrosscorr}), which decreases the
certainty with which we are able to trace the reverberated continuum
variations in the line light curve.  Since there was neither an
improvement over nor a discrepancy with past measurements, this new
result is simply added to past results as an additional measurement of
the \Hbeta-based BLR radius and $M_{\rm BH}$ in Mrk\,817.

\subsection{The BLR Radius Luminosity Relationship}
\label{S:discussRL}

To investigate the outcome of our goal to improve the calibration of
scaling relations by re-examining objects that had large measurements
uncertainties and/or that appeared as outliers on these scaling
relationships, we place our new measurements in context to the \rl\
relationship most recently calibrated by \citet{Bentz09a}.  Luminosities
were measured from the average, host-corrected continuum flux density
measured within the 5100\,\AA\ rest-frame continuum windows listed for
each object in Table \ref{tab:specobs}.  For most objects, we simply
corrected for Galactic reddening along the line of sight
\citep{Schlegel98}; however, NGC\,3227 and NGC\,3516 show evidence of
internal reddening that must be taken into account in determining the
luminosity.  \citet{Gaskell04} argue that the UV-optical continua of
AGNs are all very similar, so that the reddening can be estimated by
dividing the spectrum of a reddened AGN by the spectrum of an unreddened
AGN.  In the case of NGC\,3227, we use the value of $A_B$ determined by
\citet{Crenshaw01} by comparing the UV-optical spectrum of NGC\,3227 to
the unreddened spectrum of NGC\,4151.  For NGC\,3516, we consider two
methods for estimating the reddening, which result in consistent
estimates of $A_B$: (1) we follow the Crenshaw et al.\ method, comparing
the spectrum of NGC\,3516 again to that of NGC\,4151, which results in
$A_B = 1.72$, and (2) we use the Balmer decrement measured from the
broad components of the \Halpha\ and \Hbeta\ emission lines to estimate
a reddening of $A_B = 1.68$.  These two values are highly consistent,
and we adopt the average between the two methods of $A_B = 1.70$.  Our
measured luminosities are given in Column 9 of Table \ref{tab:results},
where the uncertainties in the luminosities are the standard deviation
in the continuum flux over the course of the campaign, except for
NGC\,4051, where the uncertainty in the distance is added in quadrature
to this \citep[see][]{Denney09b}.

The top panel of Figure \ref{fig:rlrelation} shows the \citet{Bentz09a}
\rl\ relationship, reproduced from the bottom panel of their Figure 5.
Here, we have differentiated the objects targeted for our present
campaign with solid squares, while all other objects presented by Bentz
et al.\ are open squares.  The bottom panel of Figure
\ref{fig:rlrelation} shows our current results, where the objects for
which our new measurements are either truly new (i.e., Mrk\,290) or have
become replacements for old values are shown by the solid stars, and we
no longer plot the old values.  Our additional measurements for
NGC\,5548 and Mrk\,817 are shown with the open stars, and the previous
weighted average lags and luminosities for these objects as reported by
Bentz et al. are still present in this bottom panel.  The reader should
immediately notice the increased precision and accuracy of our new and
replacement measurements, where it is important to note that we have
{\it not} determined a new fit to the data\footnote{Re-evaluating the
fit to and scatter in this relationship is outside the scope of this
paper but is planned for future work that will include all new, relevant
data \citep[see, e.g.,][]{Bentz09c}.}.  Clearly, these better
measurements emphasize the small intrinsic scatter in this relationship,
reinforcing the apparently homologous nature of AGNs, even over many
orders of magnitude in luminosity.  The results from this campaign also
support the conclusion of \citet{Peterson10} that improving this
relationship further will not come from simply obtaining more BLR radii
measurements to ``beat down'' the noise, but rather, from more reliable,
higher-precision measurements.

\subsection{Velocity-Resolved Results}

The cleanest cases of a velocity-resolved reverberation response are for
NGC\,3516, NGC\,3227, and NGC\,5548, where we see kinematic signatures
indicating apparent infall, outflow, and non-radial, or ``virialized,''
motions, respectively.  \citet{Denney09c} discuss the velocity-resolved
results for these three objects and the implications of these different
kinematic signatures in the context of our overall understanding of the
BLR and the use of BLR radii measurements for determining black hole
masses.  In addition, \citet{Denney09b} present and discuss the
marginally velocity-resolved lags shown here for NGC\,4051, and so those
results are not discussed further here.

The objects not discussed in previous publications are Mrk\,290 and
Mrk\,817.  Figure \ref{fig:reslags} shows that there is very little
variation in the reverberation lag across the full width of the Mrk\,290
line profile, indicating that any differences in the reverberation lag
across the extent of the \Hbeta-emitting region in this object were
unresolvable with the sampling rate of our campaign.  An additional
possibility for the uniform response we observed (i.e., small range in
lags and no short lags observed) could be that the highest velocity gas
seen in the wings of the mean spectrum is optically thin, and therefore
does not respond to the continuum variations.  This is supported by the
narrowness of the \Hbeta\ profile in the rms spectrum compared to that
observed in the mean spectrum.  On the other hand, based on the relative
emission-line strengths of the high-velocity wings in several AGNs,
\citet{Snedden&Gaskell07} argue against this interpretation.

At first glance, Mrk\,817 appears to show an outflow signature similar
to that of NGC\,3227, however, cross correlation between the continuum
light curve and those derived from the line flux in the first four
velocity bins actually results in lag determinations that are, though
negative, largely consistent with zero lag.  Ignoring these first bins
gives results similar to Mrk\,290, where no velocity-dependent
differences in the lags are resolved.  Taken at face value, this result
is curious.  We present binned light curves of the Mrk\,817 line profile
in Figure \ref{fig:m817lcs}, where to increase the clarity of the
discrepancy between the red and blue sides of the line for this
discussion, we have combined sets of two bins to make a total of 4 bins
instead of eight, i.e., we plot the flux from bin 1 added to that of bin
2, bin 3 added to bin 4, etc.  For completeness we also recompute the
CCFs (also shown in Fig. \ref{fig:m817lcs}) and velocity-resolved lag
measurements for these four combined bins and find results consistent
with simply taking the average of the lags of each set of two bins that
we combined, though the uncertainties in the newly measured lags are
generally smaller, particularly for the bluest and reddest bins.  Upon
inspection of the individual light curves for these bins, it becomes
apparent that the cross correlation analysis for these bins essentially
failed, not finding a strong correlation between the continuum flux
variability and that seen in the light curves of Bin 1 and Bin 2.  The
light curves show a lack of variability in the flux in these bins during
the first half of the campaign, and then a fairly monotonic rise in flux
during the second half, so the peak in the continuum flux seen near
$\sim$JD2454230 is not seen in the light curves of Bins 1 and 2, and
instead, the feature the cross correlation analysis picks up is the
trough near $\sim$JD2454282, apparently seen in the Bins 1 and 2 light
curves $\sim$8--10 days earlier.  This combination causes the cross
correlation analysis to give unreliable results.  Furthermore, no real
indication of the expected positive lag can been seen by eye, as can
with the other bins (and other objects, for that matter).  The
observations could be explained by some gas having an unresolved
velocity structure near the mean radius measured for this object and
there also being an outflowing component in the BLR of this object, so
that the blue-shifted gas is primarily along line of sight and a
resulting zero day lag is measured.  However, given that (1) the overall
variability observed in this object was small during this campaign, and
(2) the \Hbeta\ profile is very broad, leading to a small variability
signal spread over a large wavelength range, we cannot make any strong
conclusions at this time.  Future efforts will be made both to glean
further information from the velocity--delay map reconstructed from our
current data as well as to re-analyze the previous monitoring data on
this object in an attempt to search for any other indications of
velocity-resolved signatures.

Despite the differences we see in the velocity-resolved kinematics
across our sample of objects, we do not believe that there is cause for
concern for the masses derived from the mean BLR radii measured from
these reverberation lags.  Obviously, observing unresolved, virial, or
infalling gas motions certainly does not question the validity of our
assumption that the BLR motions are gravitationally dominated, but
indications of outflow may be more problematic.  However, even given
these signatures, the mean lag we measure is still consistent with lags
derived from the majority of the emission-line gas.  Besides, it is only
gas outflowing at velocities larger than the escape velocity that would
break the validity of our assumptions, and this does not seem to be the
case.  There are good observational and theoretical reasons to believe
that there are multiple components within the BLR (e.g., disk and wind
components), and the disk-wind model of \citet{Murray95}, for example,
is still able to justify the constraint of the black hole mass by the
reverberation mapping radii measurements, even with the presence of a
wind \citep[see][]{Chiang96}.

From velocity-resolved studies such as the one discussed here and in our
previous publications on this data set \citep{Denney09b,Denney09c}, it
is clear that high-cadence reverberation mapping studies are beginning
to push the envelope with respect to the amount of information we are
able to glean from data of high quality and homogeneity.  The next goal
is to attempt a reconstruction of the velocity-resolved transfer
function through the production of velocity--delay maps, with priority
placed on the objects shown here and discussed by \citet{Denney09c} that
exhibit statistically significant kinematic signatures of infall,
outflow, and virialized motions (NGC\,3516, NGC\,3227, and NGC\,5548,
respectively).  Preliminary results from this analysis show the
potential to reveal the types of structured maps that will hopefully
provide additional constraints on future models of the BLR and more
clearly reveal distinct kinematic structures responsible for the
velocity-resolved signatures we presented here.

\section{Conclusion}

We have reported the results for our complete sample of six local
Seyfert 1 galaxies that were monitored in a reverberation mapping
campaign that aimed to remeasure the BLR radius from \Hbeta\ emission in
objects that previously had poor measurements (large measurement
uncertainties and/or undersampled light curves) or that were targeted
with the aim of recovery of velocity-resolved reverberation lag signals
and/or transfer functions.  Based on the measured luminosities of our
sample over the course of our $\sim$4 month campaign, we measure \Hbeta\
lags that are in excellent agreement with the expectations of the most
recent calibration of the \rl\ relationship of \citet{Bentz09a}.

Combining these lag measurements with velocity dispersion measurements
estimated from the width of the broad \Hbeta\ emission line, we make
direct black hole mass measurements for our entire sample.  Based on a
comparison of our results with previous measurements (where available),
most of our sample constitutes results that are either entirely new
(Mrk\,290) or supersede past measurements (NGC\,3227, NGC\,3516, and
NGC\,4051).  However, for NGC\,5548 and Mrk\,817, we compared our
current mass measurements with past results and find them consistent
within the measurements uncertainties, and therefore, place these
results under the category of ``additional measurements'' for these
objects.

An additional goal of this campaign was to determine velocity-resolved
reverberation lags across the extent of the \Hbeta-emitting region of
the BLR for use in future efforts to recover velocity--delay maps to
help constrain the geometry and kinematics of the BLR.  Though the
velocity structure in some of our targets remained unresolved on
sampling-rate-limited time scales, we still found some statistically
significant and kinematically diverse velocity-resolved signatures, even
within this small sample.  We see indications of apparent infall,
outflow, and virialized motions, which, if taken at face value, would
indicate that the BLR is a complicated region that differs from object
to object.  However, given the small scatter in the \rl\ relation and
the consistency with which we are able to measure the BLR radius and
black hole mass in multiple objects across dynamical time scales (e.g.,
NGC\,5548 and Mrk\,817), it is unlikely that the steady-state dynamics
within this region are truly this diverse.  The BLR could be made up of
multiple kinematic components with possible transient features such as
winds and/or warped disks that travel through the line of sight to the
observer over dynamical timescales.  In such a scenario, evidence for
different types of kinematic signatures would arise depending on the
observer's line of sight through this region at a given time.  In order
to quantify such possibilities and fit models to the velocity-resolved
data, it is necessary to collect more velocity-resolved reverberation
mapping results for these objects, as well as others.  This remains a
goal for future observing programs, and efforts are focused on
recovering velocity--delay maps for the current sample.  Similar efforts
are being made by the LAMP consortium (M.\ Bentz, priv.\ comm.)  with the
sample presented by \citet{Bentz09c}, increasing our probability of
success for this elusive goal of reverberation mapping.

\acknowledgements We would like to thank Luis Ho for providing the
optical spectra of the \Hbeta\ and \Halpha\ regions of NGC\,3516 from
which we calculated the Balmer decrement to determine the degree of
internal reddening.  We acknowledge support for this work by the
National Science Foundation though grant AST-0604066 to The Ohio State
University.  MCB gratefully acknowledges support provided by NASA
through Hubble Fellowship grant HF--51251 awarded by the Space Telescope
Science Institute, which is operated by the Association of Universities
for Research in Astronomy, Inc., for NASA, under contract NAS 5-26555.
CMG is grateful for support by the National Science Foundation through
grants AST 03-07912 and AST 08-03883.  MV acknowledges financial support
from HST grants HST-GO-10417, HST-AR-10691, and HST-GO-10833 awarded by
the Space Telescope Science Institute, which is operated by the
Association of Universities for Research in Astronomy, Inc., for NASA,
under contract NAS5-26555.  The Dark Cosmology Centre is funded by the
Danish National Research Foundation.  VTD acknowledges the support of
the Russian Foundation for Basic Research (project no. 09-02-01136a) to
the Crimean Laboratory of the Sternberg Astronomical Institute.  SGS
acknowledges support through Grant No. 5-20 of the "Cosmomicrophysics"
program of the National Academy of Sciences of Ukraine to CrAO.  The
CrAO CCD cameras have been purchased through the US Civilian Research
and Development Foundation for the Independent States of the Former
Soviet Union (CRDF) awards UP1-2116 and UP1-2549-CR-03.  This research
has made use of the NASA/IPAC Extragalactic Database (NED) which is
operated by the Jet Propulsion Laboratory, California Institute of
Technology, under contract with the National Aeronautics and Space
Administration.


\clearpage


\clearpage


\begin{deluxetable}{lccccl}
\tablecolumns{6}
\tablewidth{0pt}
\tablecaption{Object List\label{tab:objlist}}
\tablehead{
\colhead{Objects}&\colhead{$z$}&\colhead{$\alpha_{2000}$}&
\colhead{$\delta_{2000}$}&\colhead{Host}&\colhead{$A_B$}\\
\multicolumn{2}{c}{\ }&\colhead{(hr min sec)} & 
\colhead{(\degr\ \arcmin\ \arcsec)} & \colhead{Classification} & 
\colhead{(mag)}\\
\colhead{(1)}&\colhead{(2)}&\colhead{(3)}&\colhead{(4)}&\colhead{(5)}&
\colhead{(6)}}

\startdata
Mrk\,290  & 0.02958 & 15 35 52.3 & +57 54 09 & E1 & 0.065 \\
Mrk\,817  & 0.03145 & 14 36 22.1 & +58 47 39 & SBc & 0.029 \\
NGC\,3227 & 0.00386 & 10 23 30.6 & +19 51 54 & SAB(s) pec & 0.76\tablenotemark{a} \\
NGC\,3516 & 0.00884 & 11 06 47.5 & +72 34 07 & (R)SB(s) & 1.70\tablenotemark{a} \\
NGC\,4051 & 0.00234 & 12 03 09.6 & +44 31 53 & SAB(rs)bc & 0.056 \\
NGC\,5548 & 0.01717 & 14 17 59.5 & +25 08 12 & (R')SA(s)0/a & 0.088
\enddata 

\tablenotetext{a}{Values have been adjusted to account for
additional internal reddening as described in section \ref{S:discussRL}.}

\end{deluxetable}

\begin{deluxetable}{llcccccc}
\tablecolumns{8}
\tablewidth{0pt}
\tablecaption{Spectroscopic Observations\label{tab:specobs}}
\tablehead{
\colhead{\ }&\colhead{\ }&\colhead{\ }&\colhead{Julian Dates}&\colhead{Res}&
\colhead{5100\AA\ Cont.}&\colhead{\Hbeta\ Line}&{Extraction}\\
\colhead{Objects}&\colhead{Observ.}&\colhead{$N_{\rm obs}$}&
\colhead{(-2450000)}&\colhead{(\AA)}&\colhead{Window (\AA)}&
\colhead{Limits (\AA)}&{Window (\arcsec)}\\
\colhead{(1)}&\colhead{(2)}&\colhead{(3)}&\colhead{(4)}&\colhead{(5)}&
\colhead{(6)}&\colhead{(7)}&\colhead{(8)}
}

\startdata
Mrk\,290  & MDM  & 71 & 4184--4268 & 7.6 & 5235--5265 & 4915--5086\tablenotemark{a,}\tablenotemark{b} & 5.0$\times$12.75 \\
          & CrAO & 18 & 4266--4301 & 7.5 & 5235--5265 & 4915--5086 & 3.0$\times$11.0 \\
          & DAO  & 11 & 4262--4290 & 7.9 & 5235--5265 & 4915--5086 & 3.0$\times$6.28 \\
Mrk\,817  & MDM  & 65 & 4185--4269 & 7.6 & 5245--5275 & 4900--5099 & 5.0$\times$12.75 \\
          & CrAO & 23 & 4265--4301 & 7.5 & 5245--5275 & 4900--5099 & 3.0$\times$11.0 \\
NGC\,3227 & MDM  & 75 & 4184--4268 & 7.6 & 5105--5135 & 4795--4942\tablenotemark{a,}\tablenotemark{b} & 5.0$\times$8.25 \\
NGC\,3516 & MDM  & 74 & 4184--4269 & 7.6 & 5130--5160 & 4845--4965\tablenotemark{b} & 5.0$\times$12.75 \\
          & CrAO & 19 & 4266--4300 & 7.5 & 5130--5160 & 4845--4965\tablenotemark{b} & 3.0$\times$11.0 \\
\enddata

\tablenotetext{a}{\Hbeta\ line limits were narrowed for the measurement
of the line width in the rms spectrum.  See Section \ref{S:mbh}
for details.}

\tablenotetext{b}{\Hbeta\ line limits were changed for the velocity-resolved lag investigation.  See Section \ref{S:velreslags} for details.}

\end{deluxetable}

\begin{deluxetable}{llcc}
\tablecolumns{4}
\tablewidth{0pt}
\tablecaption{Photometric Observations\label{tab:photobs}}
\tablehead{
\colhead{ }&\colhead{ }&\colhead{ }&\colhead{Julian Dates}\\
\colhead{Objects}&\colhead{Observatory}&\colhead{$N_{\rm obs}$}&
\colhead{(-2450000)}\\
\colhead{(1)}&\colhead{(2)}&\colhead{(3)}&\colhead{(4)}}

\startdata
Mrk\,290  & MAGNUM & 17 & 4200--4321  \\
          & CrAO   & 61 & 4180--4298  \\
          & UNebr  & 6  & 4199--4252  \\
Mrk\,817  & MAGNUM & 24 & 4185--4330  \\
          & CrAO   & 69 & 4180--4299  \\
NGC\,3227 & MAGNUM & 19 & 4181--4282  \\
          & CrAO   & 58 & 4180--4263  \\
          & UNebr  & 19 & 4195--4276  \\
NGC\,3516 & MAGNUM & 10 & 4190--4277  \\
          & CrAO   & 73 & 4181--4299  \\
          & UNebr  & 22 & 4195--4258  \\
NGC\,4051 & MAGNUM & 23 & 4182--4311  \\
          & CrAO   & 76 & 4180--4299  \\
          & UNebr  & 28 & 4195--4290  \\
NGC\,5548 & MAGNUM & 48 & 4182--4332  \\
          & CrAO   & 71 & 4180--4299  \\
          & UNebr  & 13 & 4198--4289 
\enddata

\end{deluxetable}			 

\begin{deluxetable}{lcccc}
\tablecolumns{5}
\tablewidth{0pt}
\tabletypesize{\small}
\tablecaption{Constant Spectral Properties\label{tab:constspecprop}}
\tablehead{
\colhead{ }&{FWHM([\oiii]\,$\lambda 5007$)\tablenotemark{a}}&{$F$([\oiii]$\lambda$5007)}&{\Hbeta$_{\rm nar}$}&{$F_{\rm Host}$}\\
\colhead{Objects}&{rest frame (km s$^{-1}$)}&($10^{-13}$\,ergs s$^{-1}$\,cm$^{-2}$)&Line Strength\tablenotemark{b}&{($10^{-15} $\,\Flamunit)}\\
\colhead{(1)}&{(2)}&{(3)}&{(4)}&{(5)}}
\startdata

Mrk\,290  & 380 & $1.91\pm0.12$ & 0.08 & 1.79\\
Mrk\,817  & 330 & $1.32\pm0.07$ & 0.08 & $1.84\pm0.17$\\
NGC\,3227 & 485 & $6.81\pm0.54$ & 0.088\tablenotemark{c} & $7.30\pm0.67$\\
NGC\,3516 & 250 & $3.35\pm0.42$ & 0.07 & $16.1\pm1.5$\\
NGC\,4051 & 190 & $3.91\pm0.12$\tablenotemark{c} & \nodata & $9.18\pm0.85$\\
NGC\,5548 & 410 & $5.58\pm0.27$\tablenotemark{d} & 0.11\tablenotemark{e} & $4.48\pm0.41$

\enddata

\tablenotetext{a}{From \citet{Whittle92}.}
\tablenotetext{b}{Ratio of narrow $F$(\Hbeta$_{\rm nar}$) to $F$([\oiii]$\lambda$5007).}
\tablenotetext{c}{From \citet{Peterson00b}.}
\tablenotetext{d}{From \citet{Peterson91}.}
\tablenotetext{e}{From \citet{Peterson04}.}

\end{deluxetable}

\begin{deluxetable}{lclclclclc}
\tablecolumns{10}
\tablewidth{0pt}
\tabletypesize{\scriptsize}
\setlength{\tabcolsep}{0.03in}
\tablecaption{$V$-band and Continuum Fluxes\label{tab:contflux}}
\tablehead{
\multicolumn{2}{c}{Mrk\,290}&\multicolumn{2}{c}{Mrk\,817}&\multicolumn{2}{c}{NGC\,3227}&\multicolumn{2}{c}{NGC\,3516}&\multicolumn{2}{c}{NGC\,5548}\\
\colhead{JD\tablenotemark{a}}&{$F_{\rm cont}$\tablenotemark{b}} &
\colhead{JD\tablenotemark{a}}&{$F_{\rm cont}$\tablenotemark{b}} &
\colhead{JD\tablenotemark{a}}&{$F_{\rm cont}$\tablenotemark{b}} &
\colhead{JD\tablenotemark{a}}&{$F_{\rm cont}$\tablenotemark{b}} &
\colhead{JD\tablenotemark{a}}&{$F_{\rm cont}$\tablenotemark{b}}}

\startdata
4180.47p & 1.083$\pm$0.015 & 4180.44p & 4.621$\pm$0.038 & 4180.28p & 3.959$\pm$0.064 &  4181.33p &   6.433$\pm$0.104&  4180.41p &   2.800$\pm$0.055\\
4181.54p & 1.070$\pm$0.015 & 4181.52p & 4.622$\pm$0.036 & 4181.32p & 3.971$\pm$0.057 &  4182.39p &   6.135$\pm$0.126&  4181.50p &   2.878$\pm$0.058\\
4184.97m & 1.102$\pm$0.047 & 4185.02g & 4.654$\pm$0.048 & 4181.90g & 3.250$\pm$0.052 &  4184.74m &   5.574$\pm$0.364&  4182.06g &   3.128$\pm$0.032\\
4185.96m & 1.109$\pm$0.047 & 4185.92m & 4.602$\pm$0.078 & 4182.36p & 3.836$\pm$0.059 &  4185.66m &   5.897$\pm$0.369&  4184.92m &   2.912$\pm$0.118\\
4186.61p & 1.102$\pm$0.033 & 4186.60p & 4.744$\pm$0.060 & 4184.68m & 3.363$\pm$0.149 &  4186.47p &   5.753$\pm$0.162&  4185.86m &   2.643$\pm$0.114\\
4186.94m & 1.194$\pm$0.048 & 4186.87m & 4.552$\pm$0.077 & 4185.61m & 3.623$\pm$0.153 &  4187.36p &   5.823$\pm$0.139&  4186.58p &   2.709$\pm$0.062\\
4187.48p & 1.184$\pm$0.021 & 4187.46p & 4.834$\pm$0.052 & 4186.45p & 3.857$\pm$0.058 &  4188.35p &   5.579$\pm$0.185&  4186.83m &   2.624$\pm$0.113\\
4187.96m & 1.242$\pm$0.049 & 4188.49p & 4.778$\pm$0.046 & 4187.35p & 3.915$\pm$0.079 &  4188.66m &   6.065$\pm$0.373&  4188.47p &   2.627$\pm$0.060\\
\enddata

\tablenotetext{a}{Julian Dates are $-2450000$ and include the following
observatory code to indicate the origin of the observation: MDM --- m,
MAGNUM --- g, CrAO spectroscopy --- c, CrAO photometry --- p, UNebr. --- n,
and DAO --- d.}

\tablenotetext{b}{Continuum fluxes are in units of $10^{-15}$ ergs
s$^{-1}$ cm$^{-2}$ \AA$^{-1}$ and represent the average continuum flux
density measured $\sim$5100\,\AA, rest-frame, from spectroscopic
observations or the photometric $V$-band flux.  Spectroscopic and
photometric fluxes were scaled to a uniform scale as described in
Section \ref{S:lightcurves}.  All fluxes have been corrected for host
starlight contamination.}

\end{deluxetable}

\begin{deluxetable}{lclclclclc}
\tablecolumns{10}
\tablewidth{0pt}
\tabletypesize{\scriptsize}
\setlength{\tabcolsep}{0.03in}
\tablecaption{\Hbeta\ Fluxes\label{tab:hbetaflux}}
\tablehead{
\multicolumn{2}{c}{Mrk\,290}&\multicolumn{2}{c}{Mrk\,817}&\multicolumn{2}{c}{NGC\,3227}&\multicolumn{2}{c}{NGC\,3516}&\multicolumn{2}{c}{NGC\,5548}\\
\colhead{JD\tablenotemark{a}}&{$F_{\Hbeta}$\tablenotemark{b}} &
\colhead{JD\tablenotemark{a}}&{$F_{\Hbeta}$\tablenotemark{b}} &
\colhead{JD\tablenotemark{a}}&{$F_{\Hbeta}$\tablenotemark{b}} &
\colhead{JD\tablenotemark{a}}&{$F_{\Hbeta}$\tablenotemark{b}} &
\colhead{JD\tablenotemark{a}}&{$F_{\Hbeta}$\tablenotemark{b}}}

\startdata
4184.97m & 2.203$\pm$0.049 & 4185.92m & 2.491$\pm$0.082 & 4184.68m & 3.559$\pm$0.103 & 4184.74m & 5.833$\pm$0.185 & 4184.92m & 2.285$\pm$0.212\\
4185.96m & 2.253$\pm$0.050 & 4186.87m &	2.515$\pm$0.083 & 4185.61m & 3.576$\pm$0.104 & 4185.66m & 6.170$\pm$0.195 & 4185.86m & 2.298$\pm$0.213\\
4186.94m & 2.230$\pm$0.049 & 4188.91m &	2.340$\pm$0.077 & 4187.61m & 3.506$\pm$0.102 & 4188.66m & 6.150$\pm$0.195 & 4186.83m & 2.227$\pm$0.207\\
4187.96m & 2.196$\pm$0.048 & 4189.86m &	2.169$\pm$0.071 & 4188.61m & 4.066$\pm$0.118 & 4189.71m & 6.130$\pm$0.193 & 4188.86m & 2.184$\pm$0.203\\
4188.95m & 2.183$\pm$0.048 & 4191.86m &	2.518$\pm$0.083 & 4190.61m & 3.690$\pm$0.107 & 4190.66m & 5.782$\pm$0.184 & 4189.81m & 1.941$\pm$0.181\\
4189.90m & 2.158$\pm$0.047 & 4192.90m &	2.450$\pm$0.080 & 4191.66m & 3.853$\pm$0.111 & 4191.71m & 5.633$\pm$0.179 & 4190.88m & 1.830$\pm$0.170\\
4190.93m & 2.251$\pm$0.050 & 4194.92m &	2.261$\pm$0.075 & 4192.61m & 3.774$\pm$0.110 & 4192.66m & 5.495$\pm$0.175 & 4191.81m & 1.730$\pm$0.161\\
4191.95m & 2.285$\pm$0.051 & 4201.90m &	2.416$\pm$0.080 & 4193.66m & 3.945$\pm$0.114 & 4193.75m & 5.534$\pm$0.176 & 4192.85m & 1.517$\pm$0.142\\
\enddata																      
																	      
\tablenotetext{a}{Julian Dates are $-2450000$ and include the same									      
observatory codes as Table \ref{tab:contflux}.}												      
																	      
\tablenotetext{b}{\Hbeta\ flux is in units of $10^{-13}$ ergs s$^{-1}$									      
cm$^{-2}$.}

\end{deluxetable}

\begin{deluxetable}{lccccc|ccccc}
\tablecolumns{11}
\tablewidth{0pt}
\tabletypesize{\small}
\tablecaption{Light Curve Statistics\label{tab:lcstats}}
\tablehead{
\colhead{ }&\multicolumn{5}{c}{Continuum Statistics}&\multicolumn{5}{c}{\Hbeta\ Line Statistics}\\
\hline
\colhead{ }&\multicolumn{2}{c}{Sampling(days)}&\colhead{Mean}&{ }&{ }&\multicolumn{2}{c}{Sampling(days)}&\colhead{Mean}&{ }&{ }\\
\colhead{Objects}&{$\langle T \rangle$}&{$T_{\rm median}$}&{Flux\tablenotemark{a}}&{$F_{\rm var}$}&{$R_{\rm max}$}&{$\langle T \rangle$}&{$T_{\rm median}$}&{Flux\tablenotemark{b}}&{$F_{\rm var}$}&{$R_{\rm max}$}\\
\colhead{(1)}&{(2)}&{(3)}&{(4)}&{(5)}&{(6)}&{(7)}&{(8)}&{(9)}&{(10)}&{(11)}}

\startdata
Mrk\,290  & 0.77 & 0.52 & 0.94 & 0.18 & $2.18\pm0.17$ & 1.18 & 1.00 & 2.09 & 0.07 & $1.32\pm0.05$\\
Mrk\,817  & 0.84 & 0.56 & 5.06 & 0.05 & $1.27\pm0.03$ & 1.33 & 1.00 & 2.41 & 0.05 & $1.29\pm0.06$\\
NGC\,3227 & 0.55 & 0.45 & 3.27 & 0.10 & $1.88\pm0.09$ & 1.13 & 1.00 & 3.99 & 0.08 & $1.48\pm0.06$\\
NGC\,3516 & 0.60 & 0.54 & 4.86 & 0.28 & $5.90\pm1.50$ & 1.26 & 1.00 & 5.54 & 0.15 & $1.94\pm0.15$\\
NGC\,4051 & 0.56 & 0.45 & 4.49 & 0.09 & $1.69\pm0.11$ & 1.08 & 1.00 & 4.67 & 0.07 & $1.39\pm0.07$\\
NGC\,5548 & 0.70 & 0.48 & 2.29 & 0.11 & $1.71\pm0.06$ & 1.09 & 1.00 & 1.20 & 0.26 & $3.74\pm0.49$ 
\enddata

\tablenotetext{a}{Fluxes are the same units as Table \ref{tab:contflux}.}
\tablenotetext{b}{Fluxes are the same units as Table \ref{tab:hbetaflux}.}

\end{deluxetable}

\begin{deluxetable}{lcccccccc}
\tablecolumns{9}
\tablewidth{0pt}
\tabletypesize{\small}
\tablecaption{Rest Frame Lags, Line Widths, Black Hole Masses, and Luminosities\label{tab:results}}
\tablehead{
\colhead{ }&{ }&{$\tau_{\rm cent}$}&{$\tau_{\rm peak}$}&{$\sigma_{\rm line}$}&
\colhead{FWHM}&{$M_{\rm vir}$}&{$M_{\rm BH}$\tablenotemark{a}}&{log\,$L_{5100}$}\\
\colhead{Objects}&{$r_{\rm max}$}&{(days)}&{(days)}&{(km/s)}&{(km/s)}&
\colhead{($\times 10^{6}M_{\odot}$)}&{($\times 10^{6}M_{\odot}$)}&{(ergs s$^{-1}$)}\\
\colhead{(1)}&{(2)}&{(3)}&{(4)}&{(5)}&{(6)}&{(7)}&{(8)}&{(9)}}
\startdata
Mrk\,290 & 0.632 & $8.72^{+1.21}_{-1.02}$ & $9.2^{+1.5}_{-1.4}$ & $1609 \pm 47 $ & $4270 \pm 157$ & $4.42^{+0.67}_{-0.67}$ & $24.3^{+3.7}_{-3.7}$ & $43.00^{+0.08}_{-0.08}$\\
Mrk\,817\tablenotemark{b}  & 0.614 & $14.04^{+3.41}_{-3.47}$ & $16.0^{+3.9}_{-5.3}$ & $2025 \pm 5$ & $5627 \pm 30$ & $11.3^{+2.7}_{-2.8}$ & $43.3^{+10.5}_{-10.7}$ & $43.78^{+0.02}_{-0.02}$ \\
NGC\,3227 & 0.547 & $3.75^{+0.76}_{-0.82}$ & $2.99^{+2.00}_{-1.00}$ & $1376 \pm 44$ & $3578 \pm 83$ & $1.39^{+0.29}_{-0.31}$ & $7.63^{+1.62}_{-1.72}$ & $42.11^{+0.04}_{-0.04}$\\
NGC\,3516 & 0.894 & $11.68^{+1.02}_{-1.53}$ & $7.43^{+1.99}_{-0.99}$ & $1591 \pm 10$ & $5175 \pm 96$ & $5.76^{+0.51}_{-0.76}$ & $31.7^{+2.8}_{-4.2}$ & $43.17^{+0.15}_{-0.15}$\\
NGC\,4051 & 0.583 & $1.87^{+0.54}_{-0.50}$ & $2.60^{+0.79}_{-1.40}$ & $927 \pm 64$ & $1034 \pm 41$ & $0.31^{+0.10}_{-0.09}$ & $1.73^{+0.55}_{-0.52}$ & $41.82^{+0.10}_{-0.36}$\\
\enddata

\tablenotetext{a}{Using \citet{Onken04} calibration (except Mrk\,817,
see below).}

\tablenotetext{b}{The weak and poorly defined, triple-peaked profile of
the \Hbeta\ emission in the rms spectrum necessitated the use of the
line width measured from the mean spectrum for Mrk\,817 (Columns 5 and
6) and a black hole mass (Column 8) calculated with the scale factor
determined by \citet{Collin06} for the use of this line width
measurement, $f=3.85$, instead of the standard \citet{Onken04} value of
$f=5.5$ that was used for all other objects.}

\end{deluxetable}

\clearpage


\begin{figure}
\figurenum{2}
\epsscale{1}
\plotone{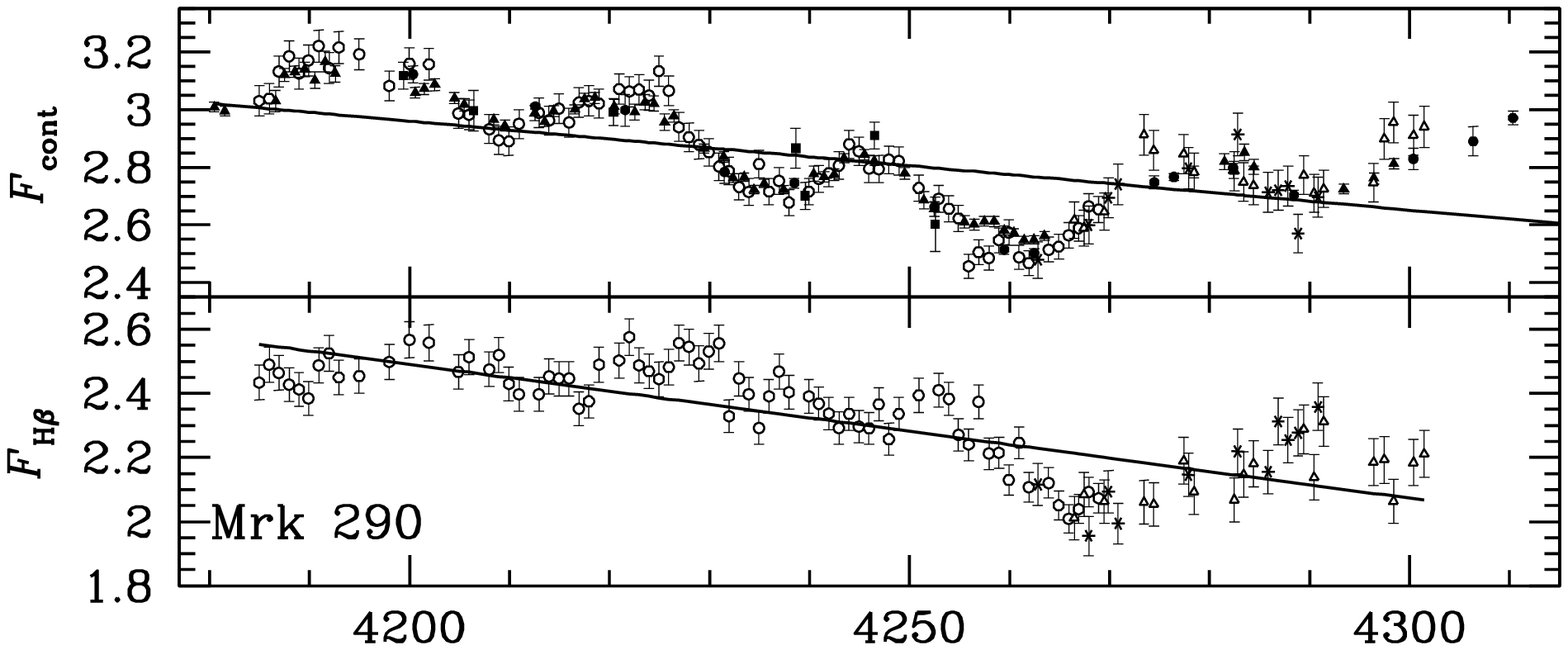}
\plotone{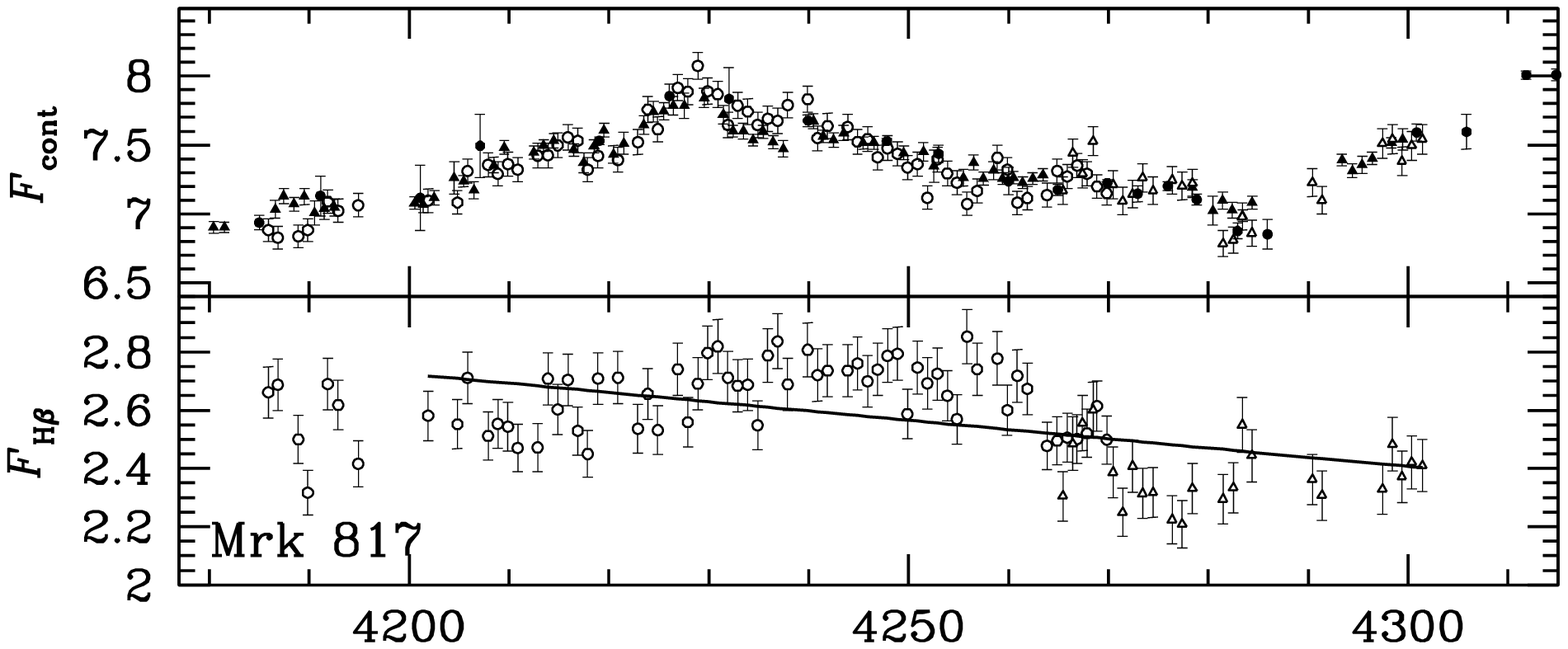}
\plotone{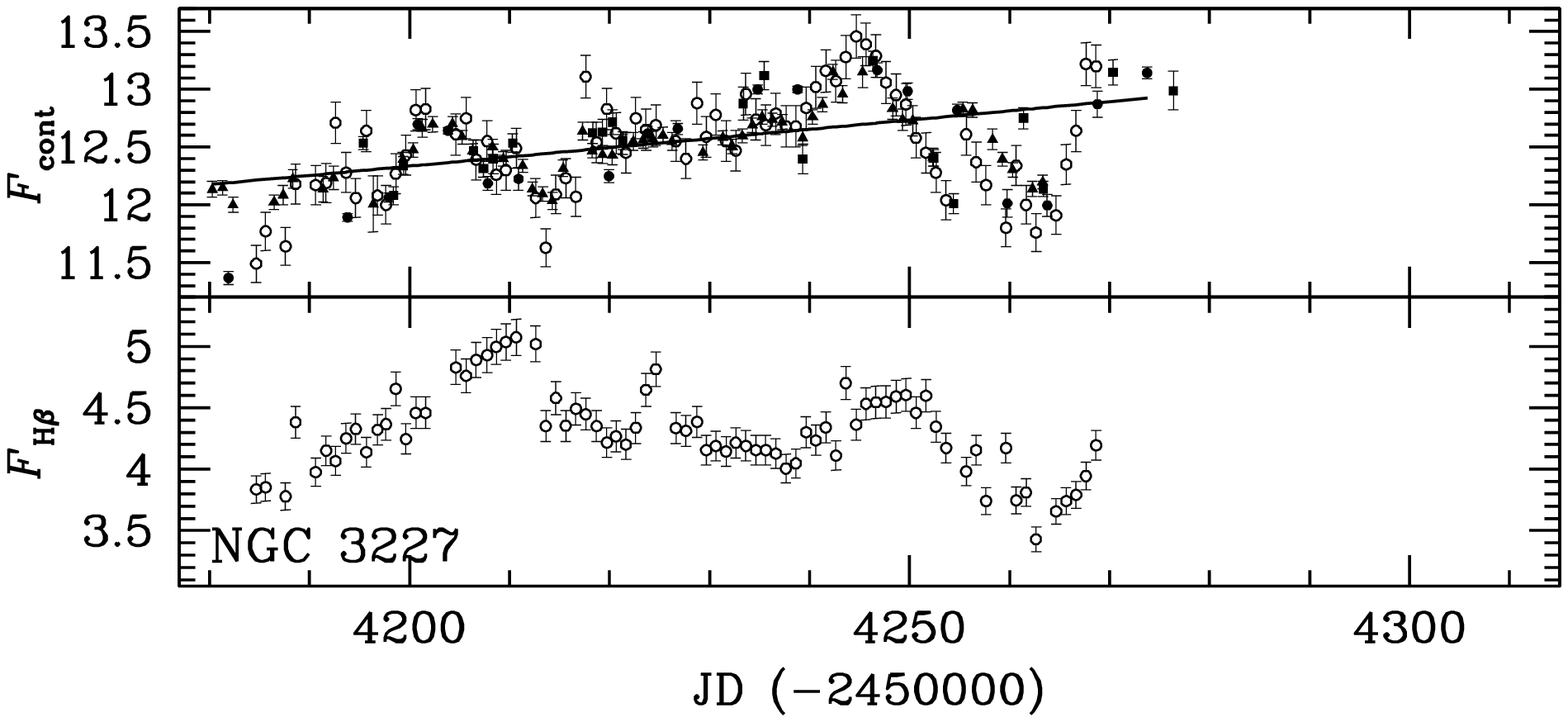}

\caption{Light curves showing complete set of observations from all
sources for all objects.  {\it Top:} The 5100\,\AA\ continuum flux in
units of $10^{-15}$ ergs s$^{-1}$ cm$^{-2}$ \AA$^{-1}$. {\it Bottom:}
H$\beta$ $\lambda$4861 line flux in units of $10^{-13}$ ergs s$^{-1}$
cm$^{-2}$.  Observations from different sources are as follows: CrAO
photometry --- solid triangles, MAGNUM photometry --- solid circles,
UNebr. photometry --- solid squares, MDM spectroscopy --- open circles,
CrAO spectroscopy --- open triangles, and DAO spectroscopy ---
asterisks. The solid lines show linear, secular-variation detrending fits
to the light curves.}

\label{fig:lcsep}
\end{figure}

\begin{figure}\ContinuedFloat
\figurenum{2}
\ContinuedFloat
\epsscale{1}
\plotone{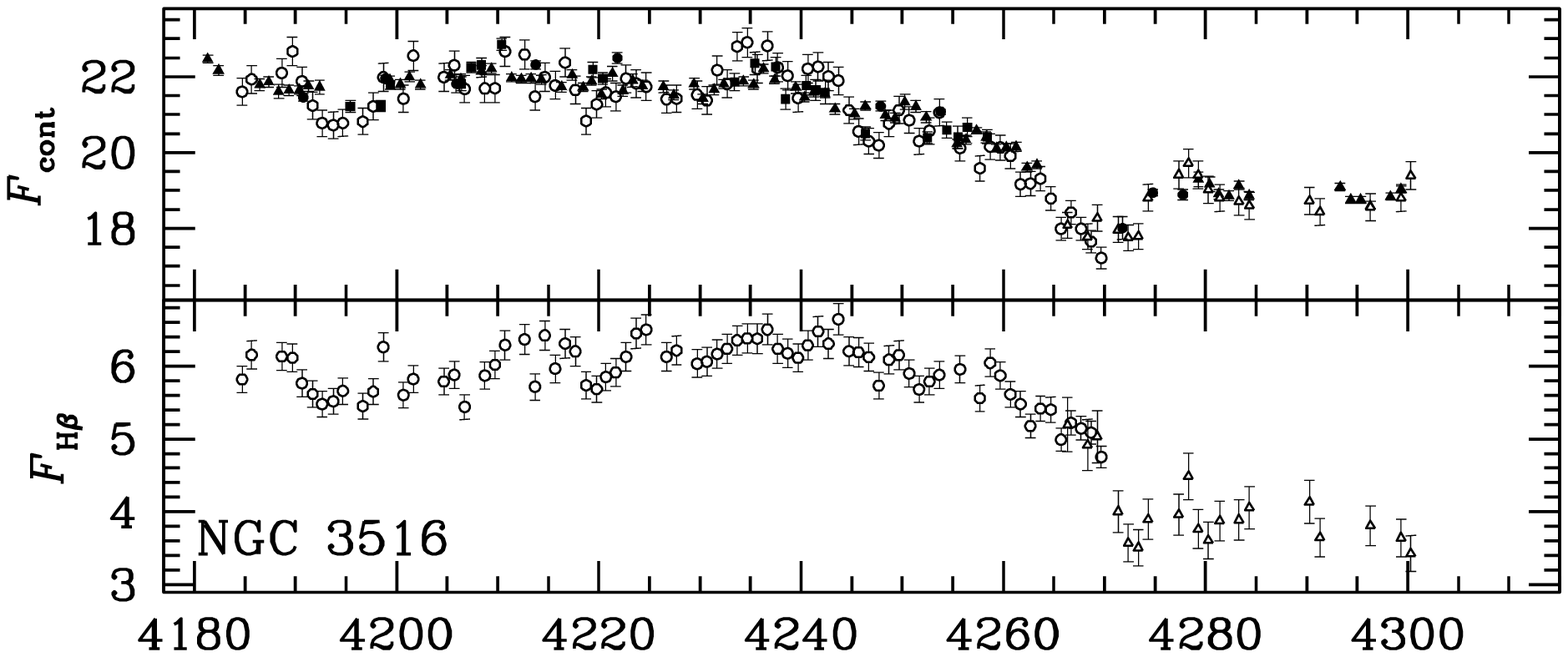}
\plotone{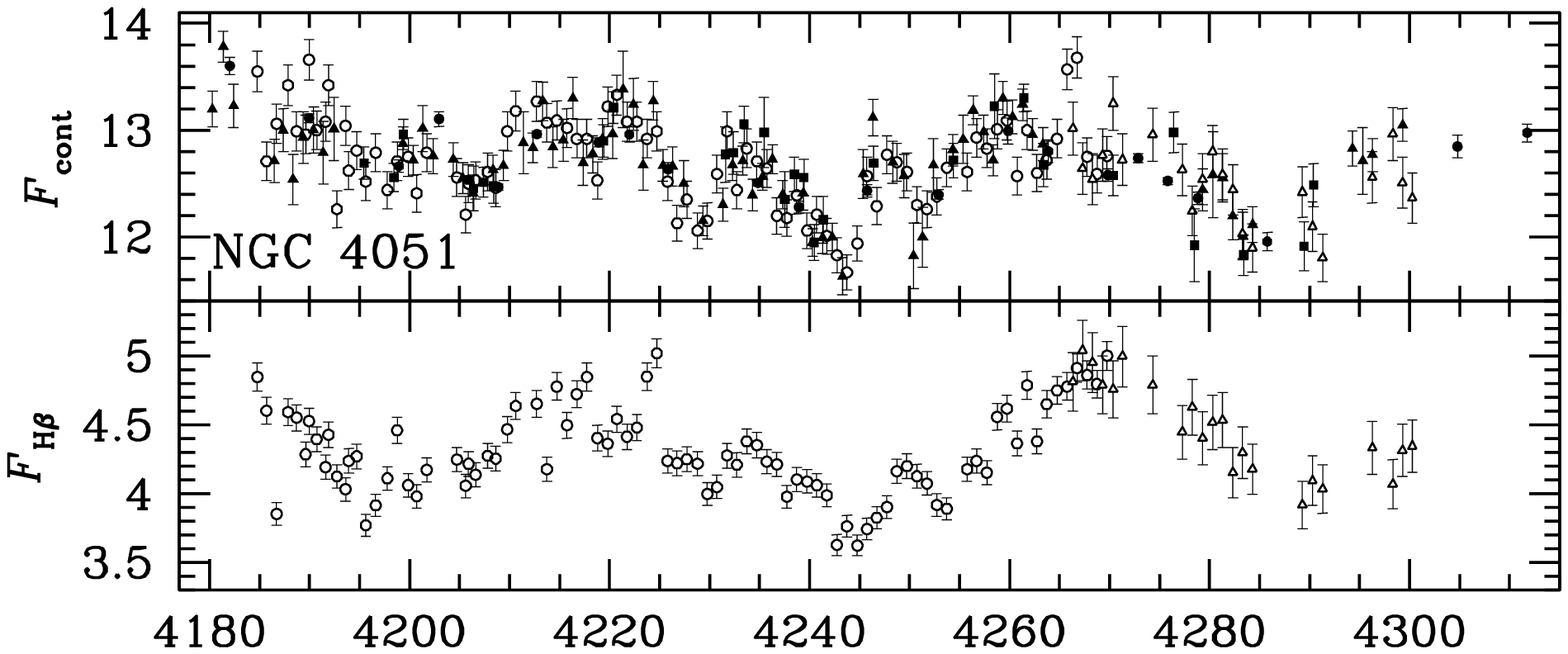}
\plotone{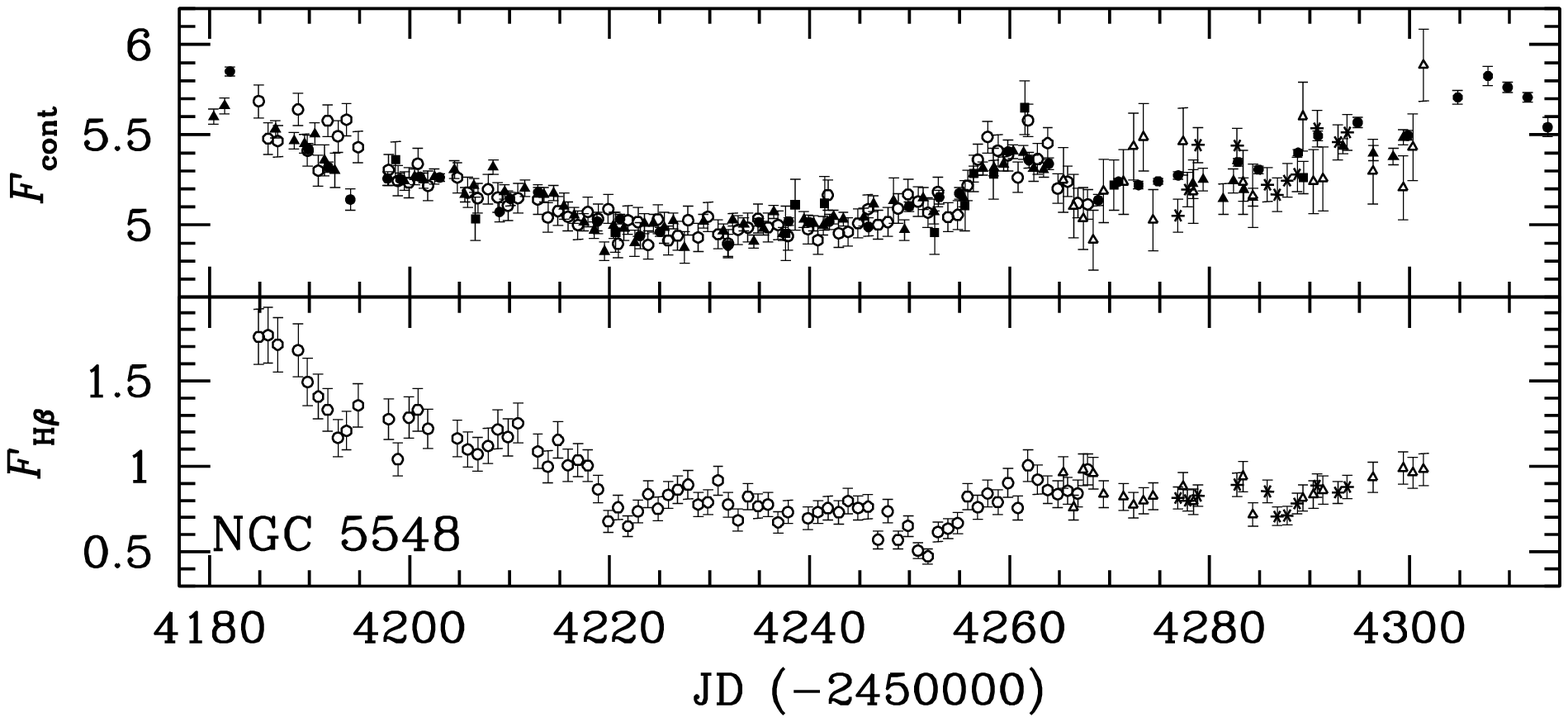}

\caption{{\it Continued.}}

\end{figure}

\begin{figure}
\figurenum{3}
\epsscale{1}
\plotone{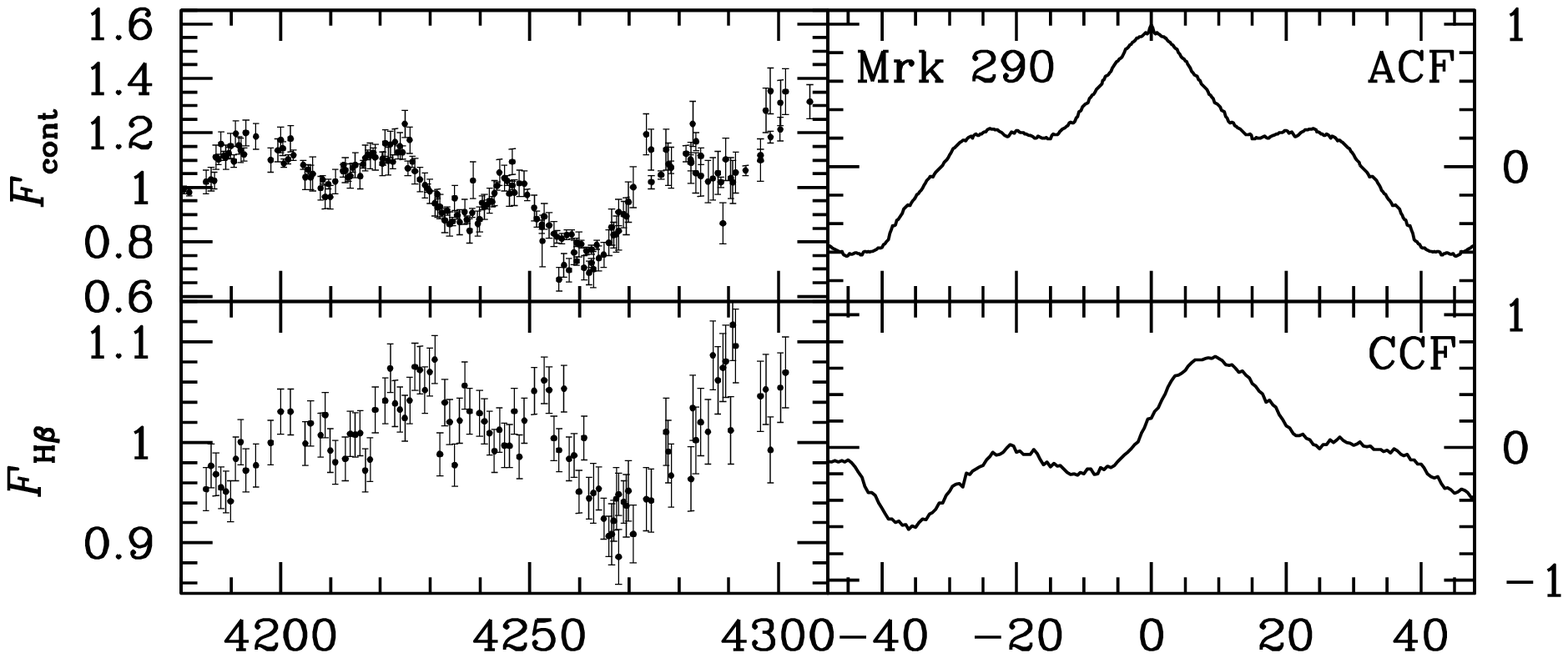}
\plotone{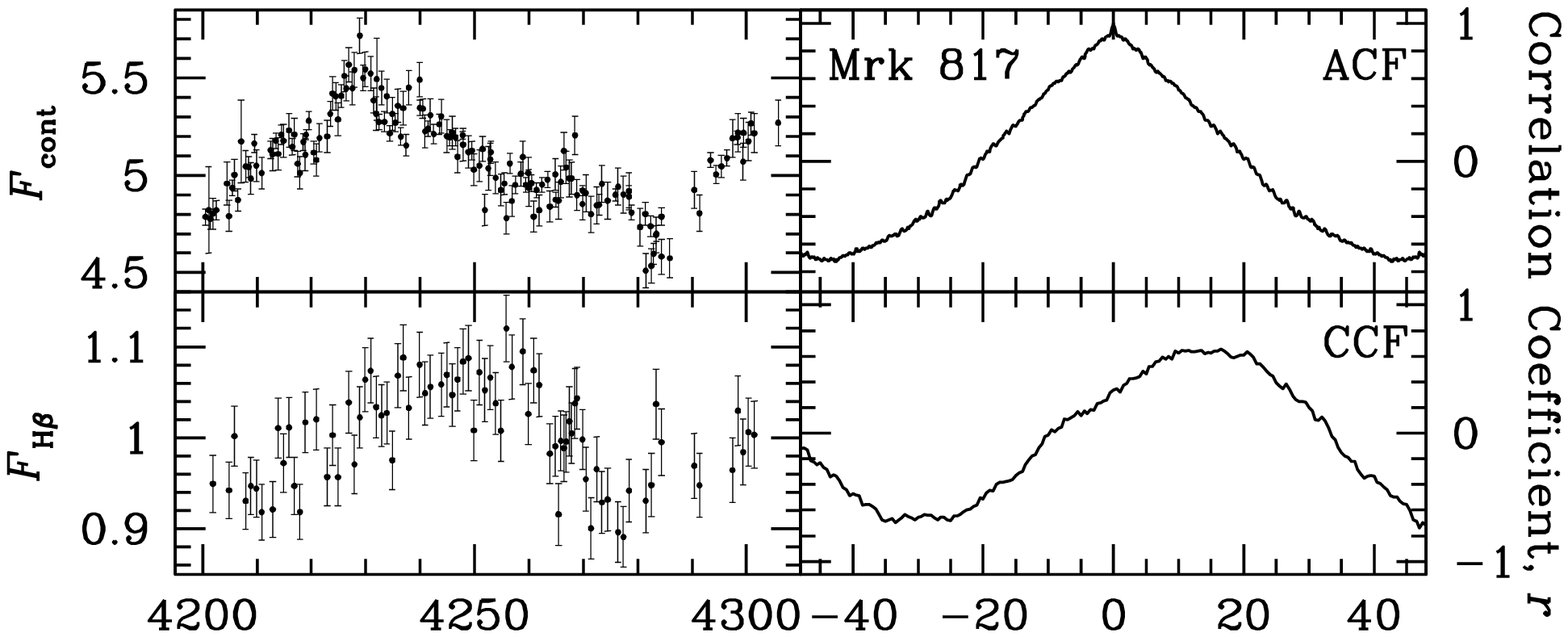}
\plotone{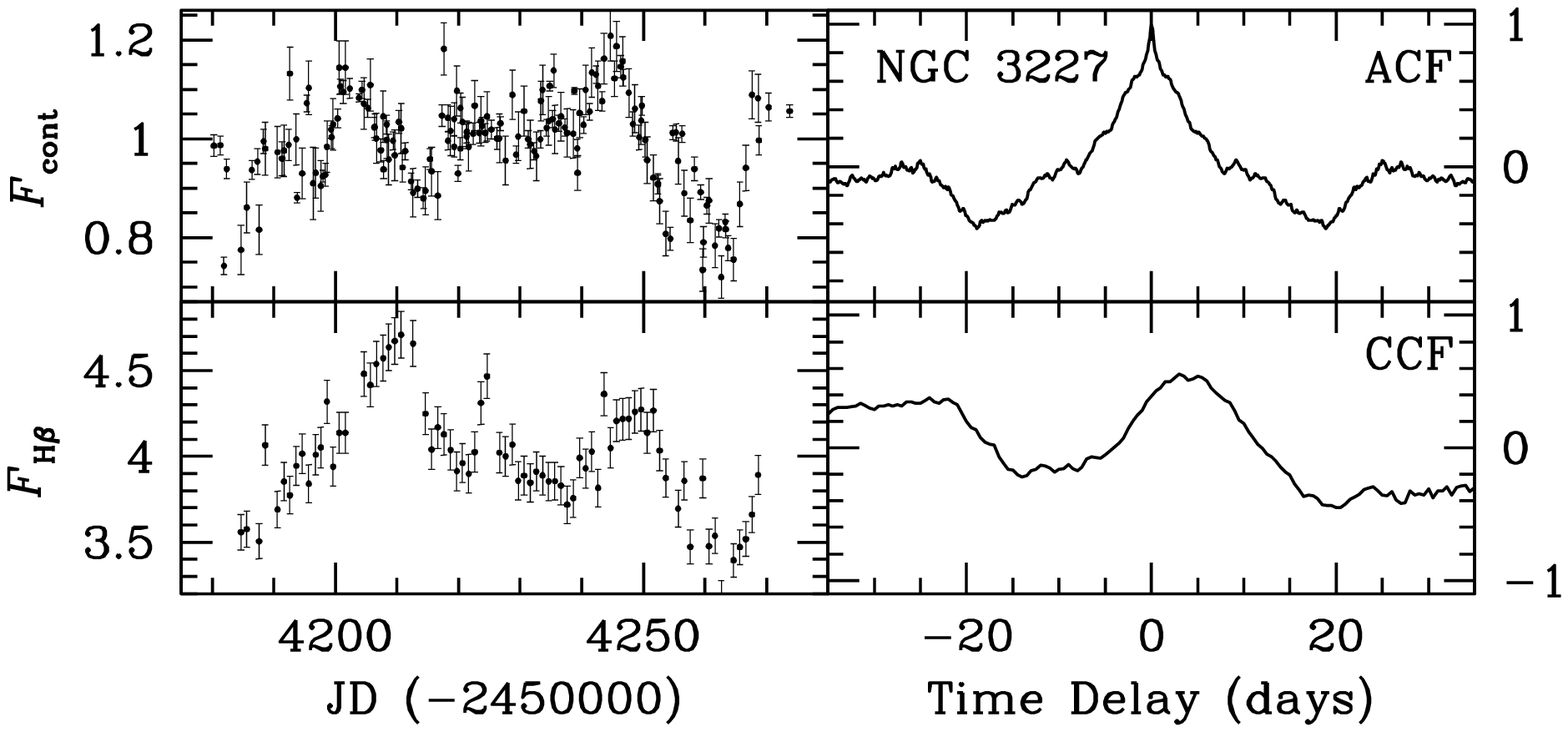}

\caption{{\it Left panels:} Merged and detrended (where applicable)
continuum (top) and \Hbeta\ (bottom) light curves used for cross
correlation analysis.  Units are the same as Tables \ref{tab:contflux}
and \ref{tab:hbetaflux}, but the flux scale of each detrended light
curve is arbitrary.  {\it Right panels:} Cross-correlation functions for
the light curves.  Each top panel shows the autocorrelation function of
each continuum light curve, and the bottom panels show the
cross-correlation function of \Hbeta\ with the continuum.}

\label{fig:lccrosscorr}
\end{figure}

\begin{figure}\ContinuedFloat
\figurenum{3}
\ContinuedFloat
\epsscale{1}
\plotone{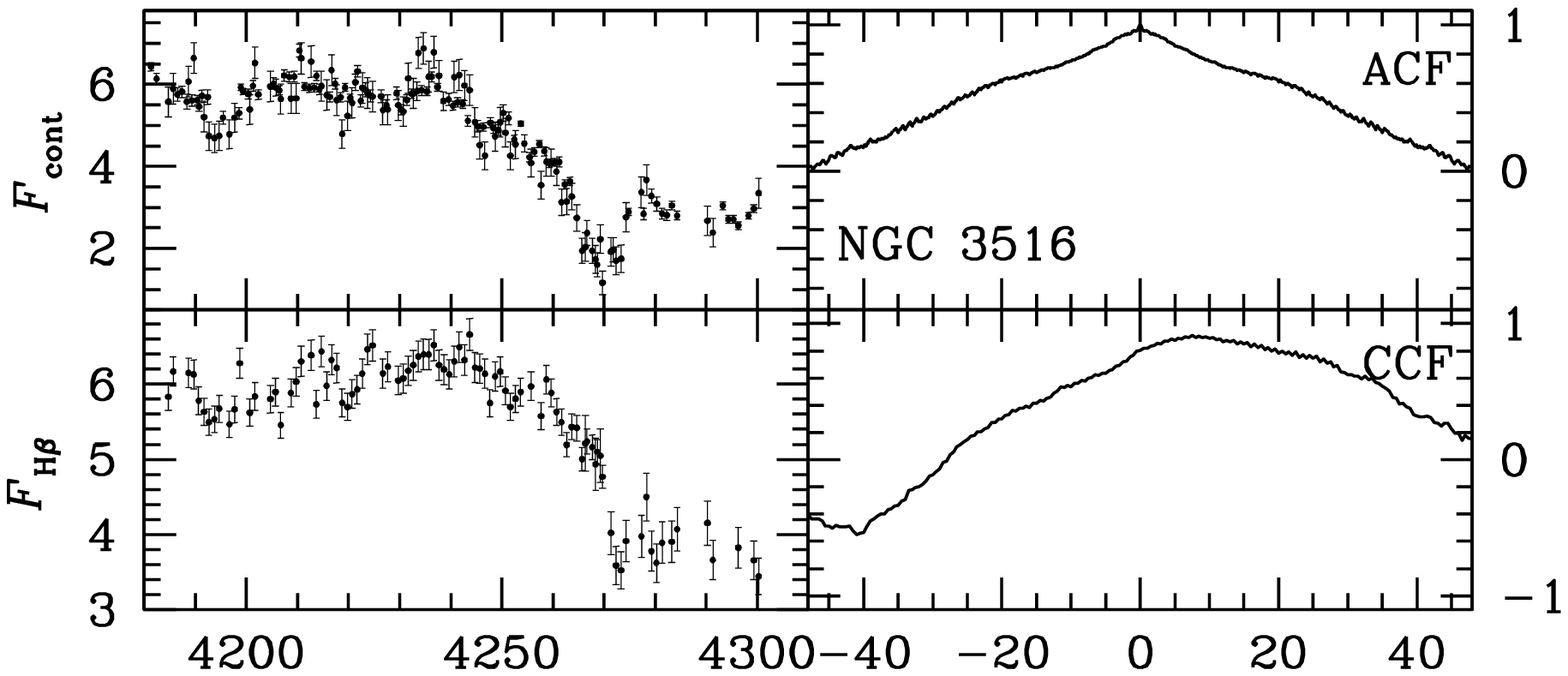}
\plotone{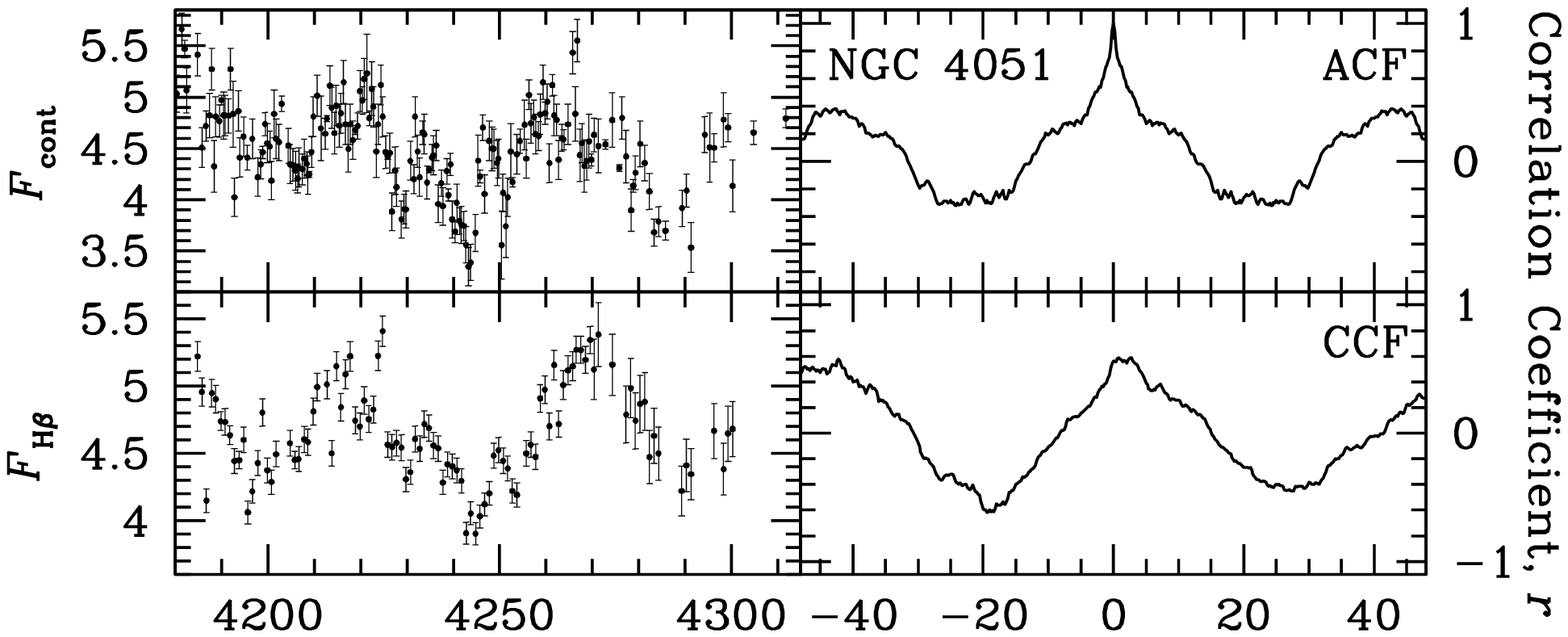}
\plotone{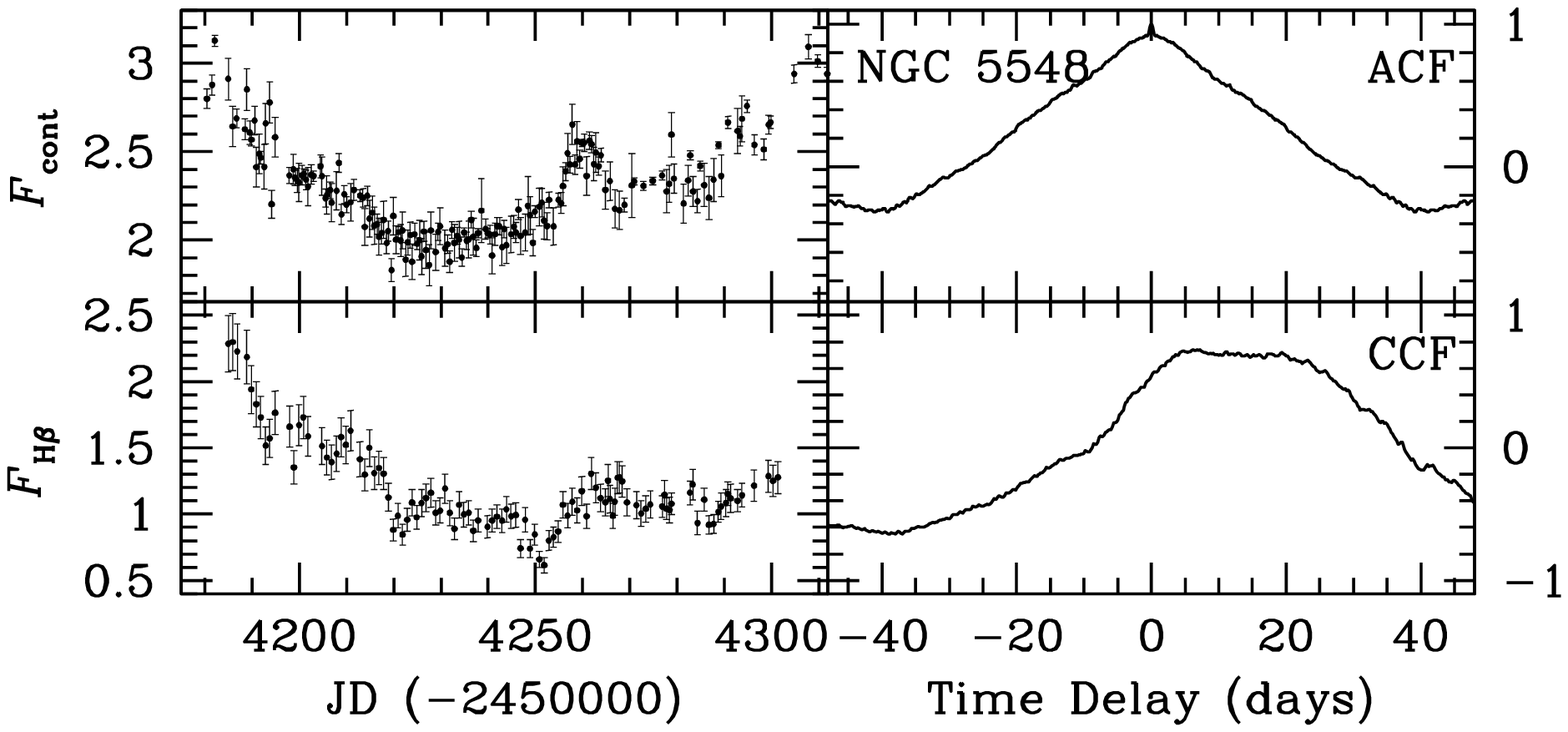}

\caption{{\it Continued.}}

\end{figure}

\begin{figure}
\figurenum{4}
\epsscale{1}
\plotone{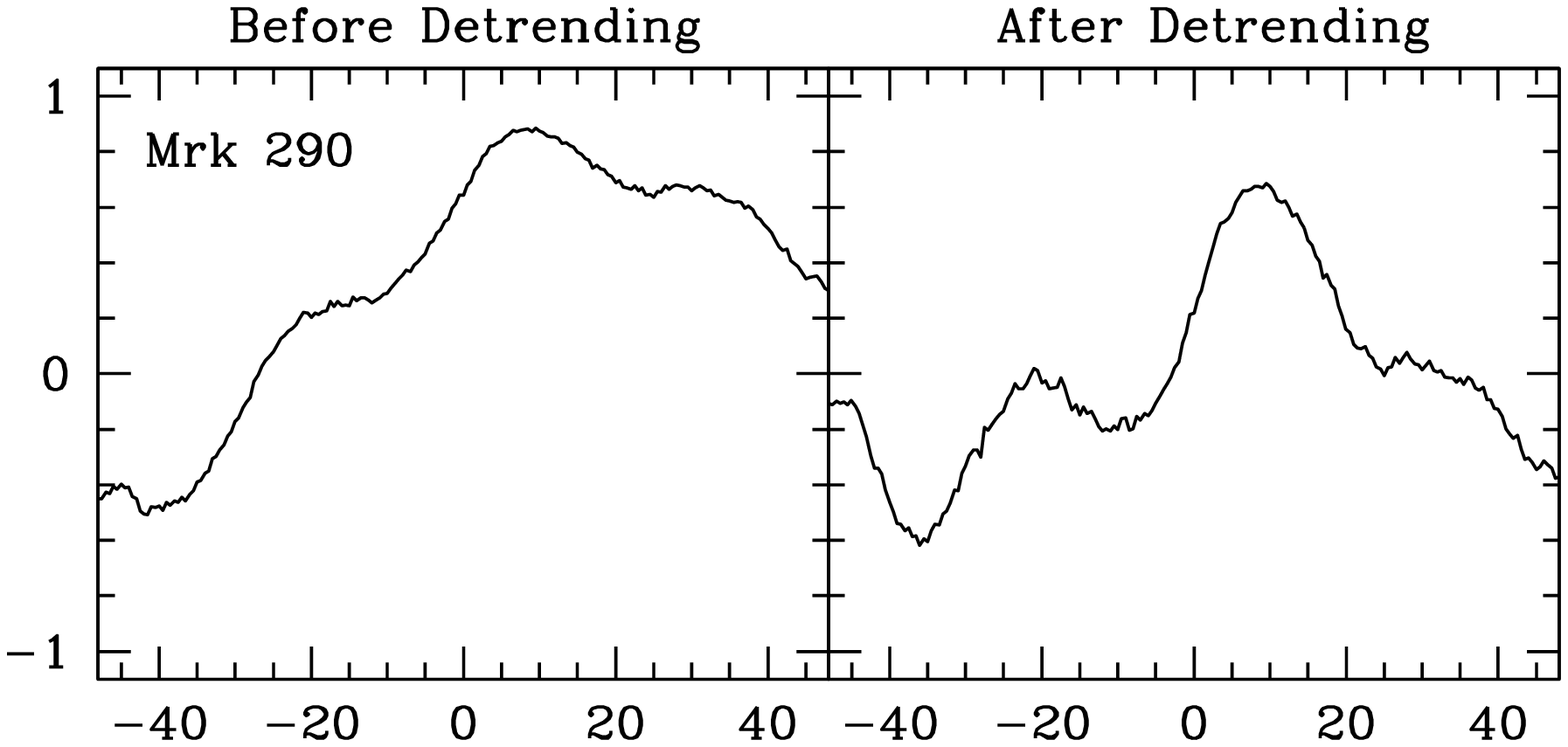}
\plotone{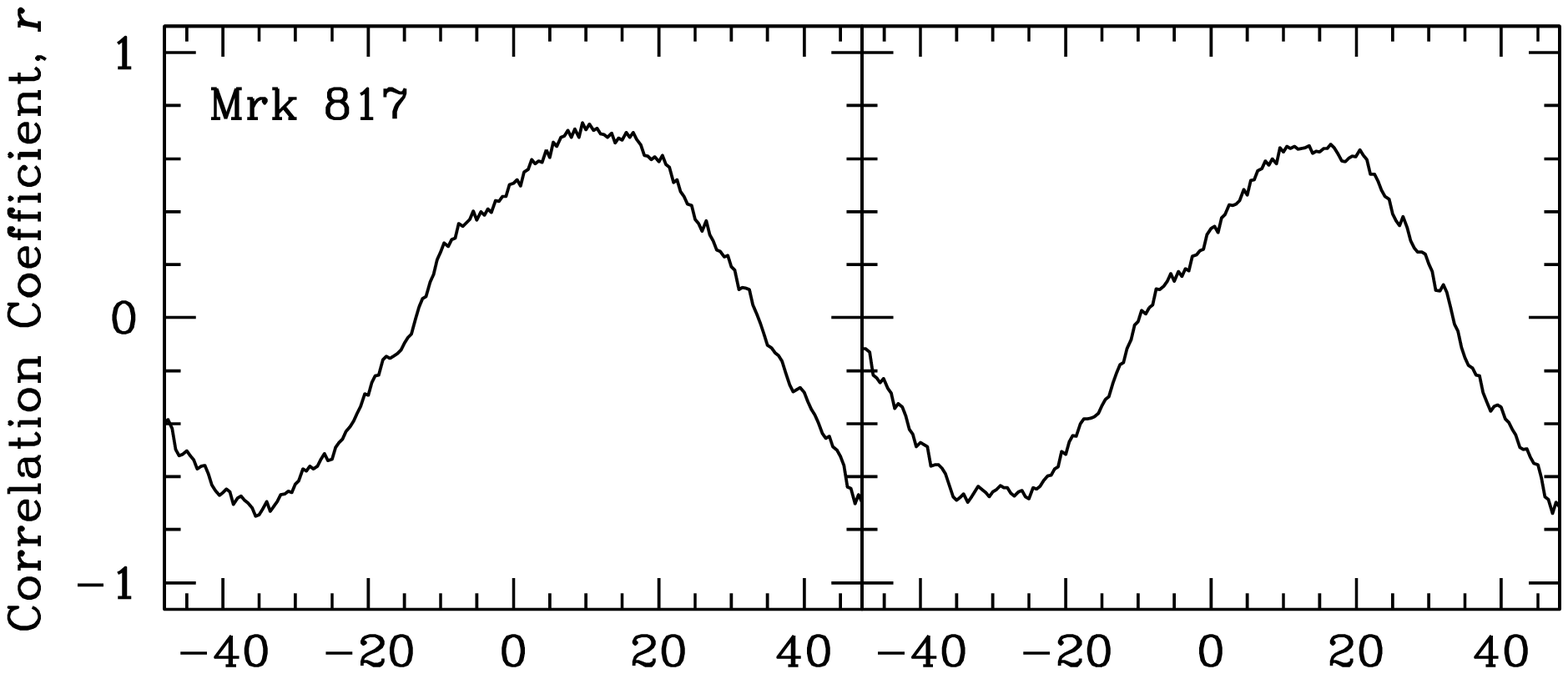}
\plotone{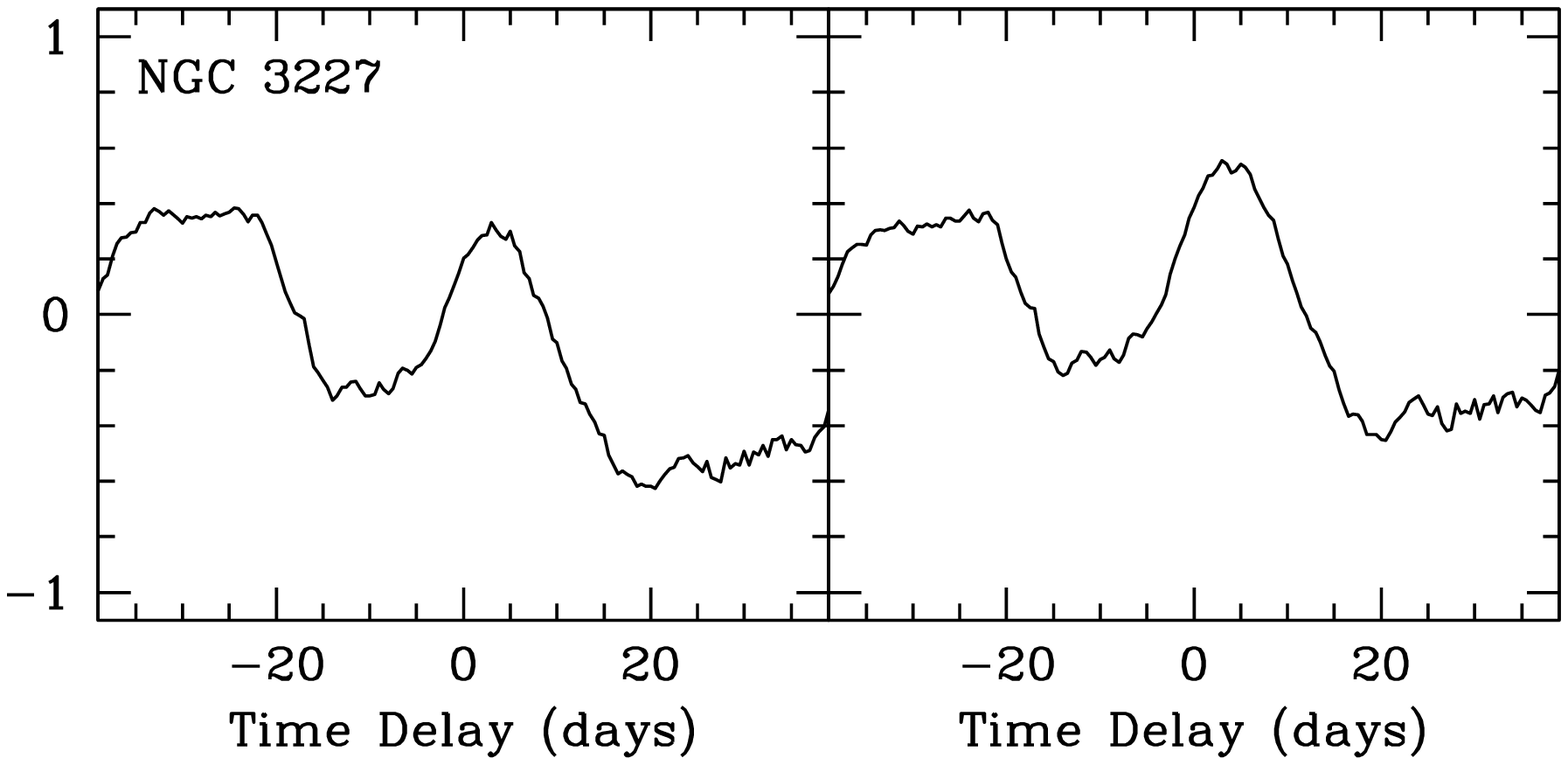}

\caption{CCFs before (left) and after (right) detrending selected light
curves of Mrk\,290 (top), Mrk\,817 (middle), and NGC\,3227 (bottom).
See Section \ref{S:lagresults} for details.}

\label{fig:detrCCFs}
\end{figure}

\begin{figure}
\figurenum{5}
\epsscale{1}
\plotone{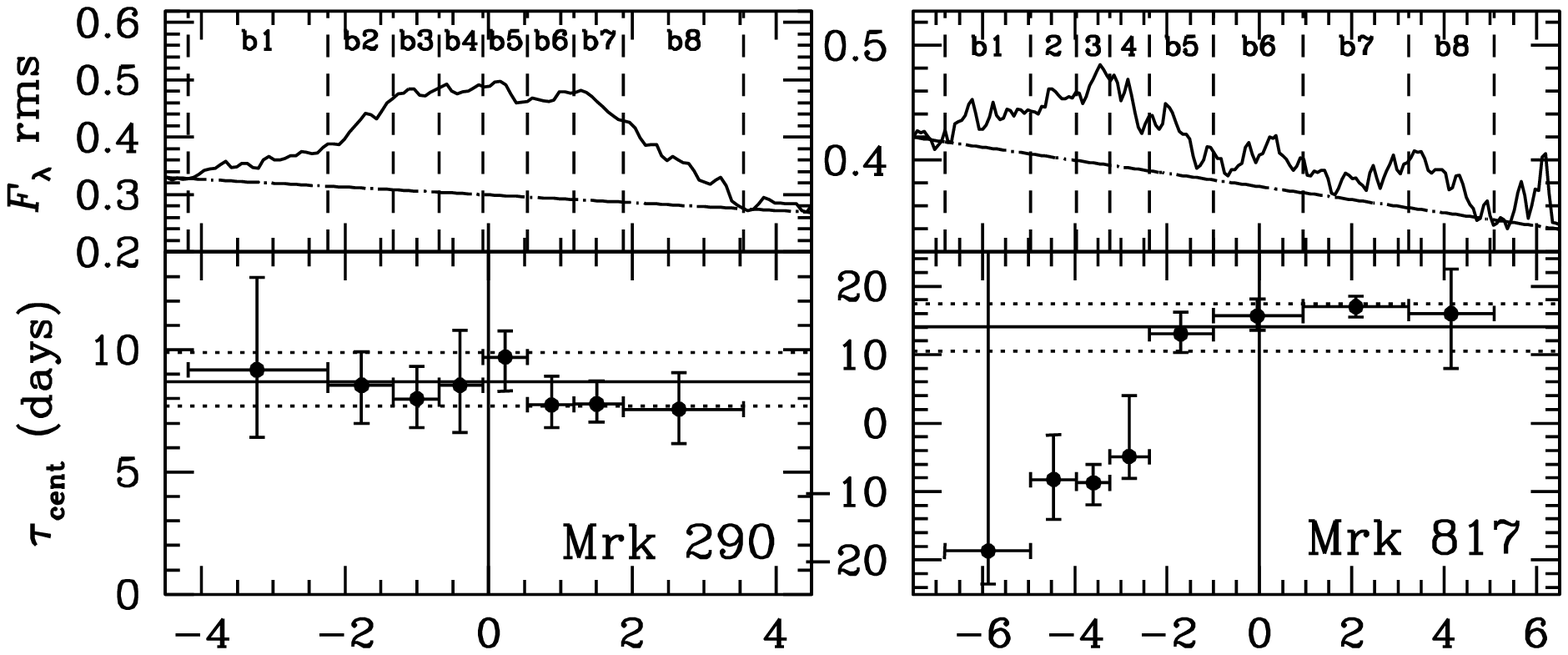}
\plotone{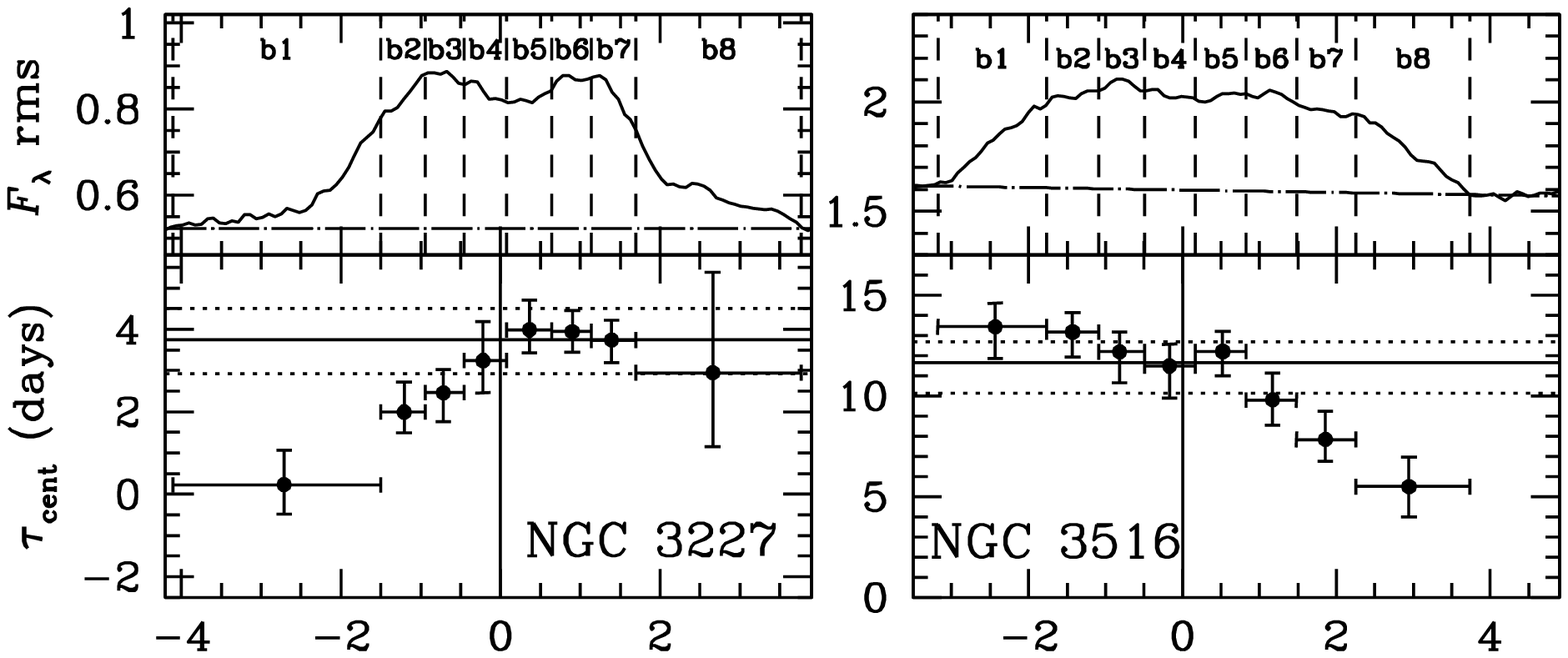}
\plotone{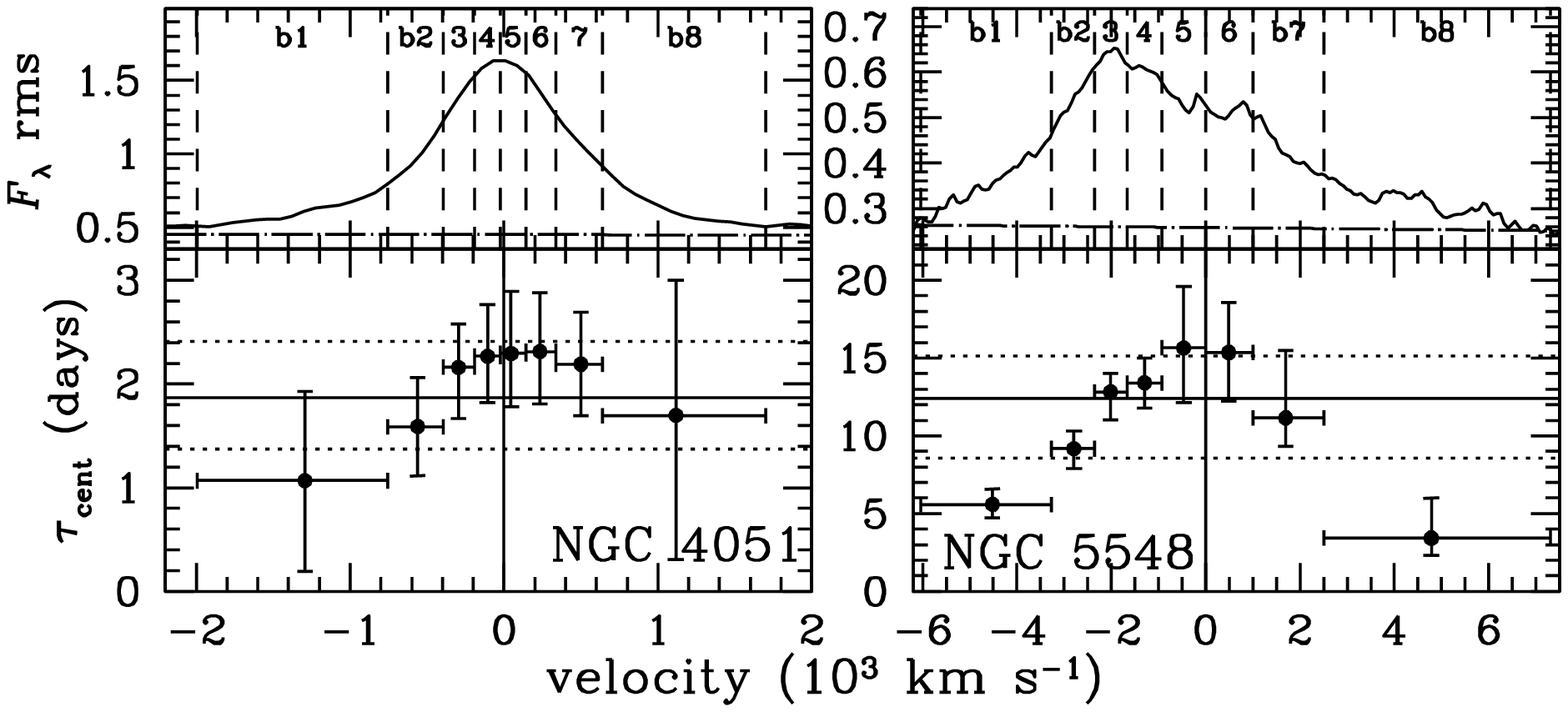}

\caption{{\it Top panels:} \Hbeta\ rms spectral profile of each object broken
into bins of equal flux (numbered and separated by dashed lines) with
the linearly-fit continuum level shown (dotted-dashed line).  Flux units
are the same as in Fig.\ \ref{fig:meanrms}.  {\it Bottom panels:}
Velocity-resolved time-delay measurements.  Time delay measurements and
errors are determined similarly to those for the mean BLR lag, and error
bars in the velocity direction show the bin size.  The horizontal solid
and dotted lines show the mean BLR centroid lag and associated errors,
calculated in Section \ref{S:lagresults}.}

\label{fig:reslags}
\end{figure}

\begin{figure}
\figurenum{6}
\epsscale{1}
\plotone{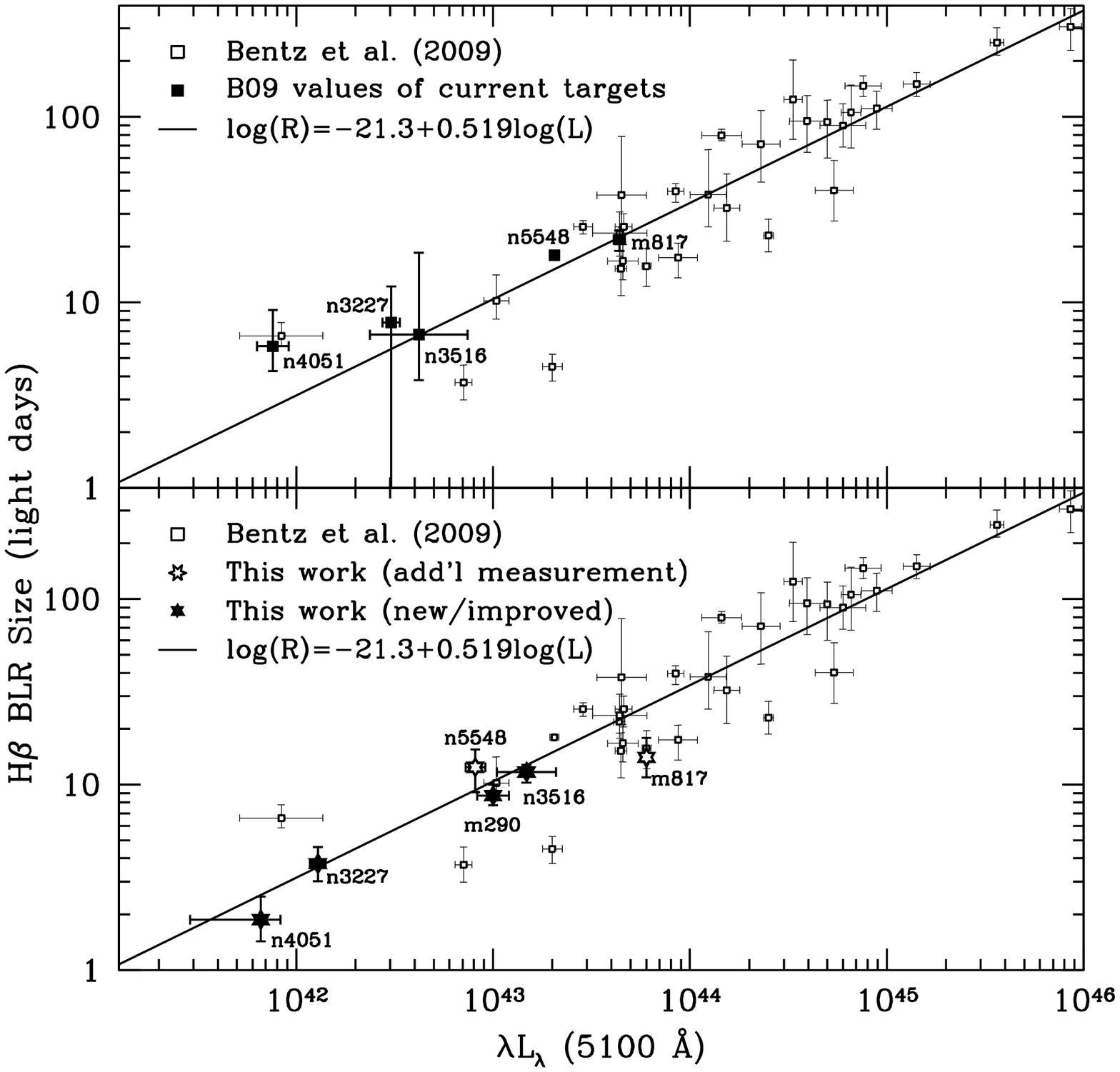}

\caption{{\it Top:} Most recently calibrated $R_{\rm BLR}$--$L$ relation
\citep[][solid line]{Bentz09a}. The closed points show the location of
our targets, and open points show all other objects used by Bentz et al.
{\it Bottom:} Same as top but with our new results displayed.  Solid
stars show new objects or improvements upon past results which replace
solid points of NGC\,4051, NGC\,3227, NGC\,3516, and Mrk\,290 in top
panel, and open points show results for NGC\,5548 and Mrk\,817, which
serve as additional measurements for these objects but do not replace
previous measurements.  Note that we keep the same calibration of the
relationship as determined by Bentz. et al.; no new fit has been
calculated with our new results.}

\label{fig:rlrelation} 
\end{figure} 

\begin{figure}
\figurenum{7}
\epsscale{1}
\plotone{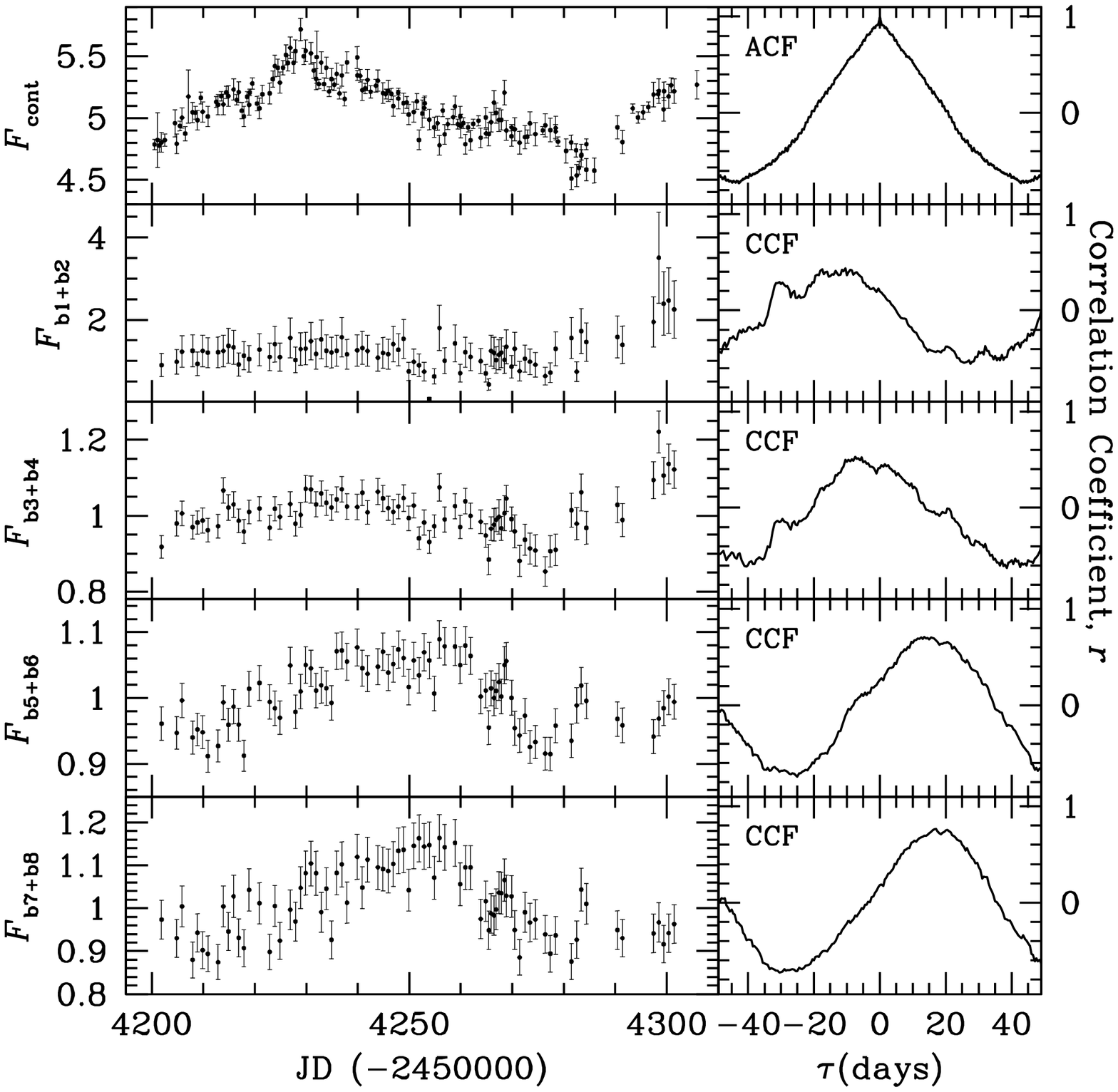}

\caption{{\it Left panels:} Continuum (top) and linearly detrended
\Hbeta\ light curves of Mrk\,817 from four equal flux bins.  Units are
the same as Tables \ref{tab:contflux} and \ref{tab:hbetaflux}.  {\it
Right panels:} Cross-correlation functions for the light curves.  The
top panel shows the autocorrelation function of the continuum light
curve, and the lower panels show the cross-correlation function of each
\Hbeta\ bin with the continuum.}

\label{fig:m817lcs}
\end{figure}

\end{document}